\documentclass[a4paper,11pt]{article}
\pdfoutput=1
\usepackage{graphicx}
\usepackage
 {
   amsmath,amssymb,tabularx
 }
\usepackage[usenames,dvipsnames]{xcolor}
\usepackage{hyperref}
\usepackage{caption,subcaption,cite}
\usepackage{geometry}
\usepackage{authblk}
\usepackage[titletoc,title]{appendix}

    \definecolor{X575}{rgb}{0.05, 0.7, 0.05}

 \newcommand{\nn}{\nonumber}

\title{Constraints on CP-violating gauge-Higgs operators}
\author{ Siddharth Dwivedi$^1$\footnote{E-mail: siddharthdwivedi@hri.res.in}, Dilip Kumar Ghosh$^2$\footnote{E-mail: tpdkg@iacs.res.in},
         Biswarup Mukhopadhyaya$^1$\footnote{E-mail: biswarup@hri.res.in} 
        and Ambresh Shivaji$^1$\footnote{E-mail: ambreshkshivaji@hri.res.in}}
  \affil{$^1$\textit{Regional Centre for Accelerator-based Particle Physics,}
  \textit{Harish-Chandra Research Institute, Chhatnag Road, Jhunsi,} \\
  \textit{Allahabad - 211019, India}}
 \affil{$^2$\textit{Department of Theoretical Physics,}
\\
 \textit{Indian Association for the Cultivation of Science,}
\\
 \textit{2A \& 2B Raja S.C. Mullick Road,}\\

 \textit{Kolkata - 700032, India}}

\begin{document}

\maketitle

\begin{abstract}\noindent

We consider the most general set of $SU(2) \times U(1)$ invariant CP-violating operators of
dimension six, which contribute to $VVh$ interactions ($V = W, Z, \gamma$). Our aim
is to constrain any CP-violating new physics above the electroweak scale via
the effective couplings that arise when such physics is integrated out. For this purpose,
we use, in turn, electroweak precision data, global fits of Higgs data at the Large Hadron Collider
and the electric dipole moments of the neutron
and the electron. We thus impose constraints mainly on two-parameter and three-parameter spaces. 
We find that the constraints from the electroweak precision data are the weakest.
 Among the existing Higgs search channels, considerable constraints come from the 
diphoton signal strength. We note that potential contribution to $h \rightarrow \gamma Z$ may in principle 
be a useful constraining factor, but it can be utilized only in the high energy run.
The contributions to  electric dipole moments mostly lead to the strongest constraints, 
though somewhat fine-tuned combinations of more than one parameter with large magnitudes are allowed. 
 We also discuss constraints on gauge boson trilinear couplings which depend on the parameters of the 
CP-violating operators .

\end{abstract}

\vfill
\begin{flushright}
HRI-RECAPP-2015-009
\end{flushright}

\newpage

\section{Introduction}

Although the discovery of ``a Higgs-like boson'' at the Large Hadron Collider (LHC)
has been a refreshing  development~\cite{Aad:2012tfa,Chatrchyan:2012ufa}, there is no clear signal yet for physics
beyond the standard model (SM). It is therefore natural that physicists are trying
to wring the last drop out of the Higgs sector itself, in attempts to read fingerprints
of new physics. 

One approach is to examine all available data in terms of specific new
models, such as supersymmetry or just additional Higgs doublets. In
the other approach, one can take a model-independent stance,
parametrize possible modifications of the interaction terms of the
Higgs with pairs of SM particles, and examine them in the light of the
available data. Such modifications can again be of two types.  In the
first category, they are just multiplicative modifications of the
coupling strengths, the Lorentz structures remaining the same as in
the SM.  Constraints on such modifications have already been derived
from the available Higgs data~\cite{Baak:2014ora,Einhorn:2013tja,Banerjee:2012xc,Giardino:2012ww,Klute:2012pu}.  In the second class, one considers
additional operators with new Lorentz structures satisfying
all symmetries of the SM~\cite{Buchmuller:1985jz,Grzadkowski:2010es,Masso:2014xra,Falkowski:2015fla,
Azatov:2012bz,Espinosa:2012ir,Contino:2013kra,Maltoni:2013sma}. Gauge invariance of such operators in their original forms
may be expected, since they are obtained by integrating out new physics that is just above 
the reach of the present round of experiments. Sets of such
higher-dimensional operators contributing to the effective coupling of
the Higgs to, say a pair of electroweak vector bosons have been
studied extensively. Here it makes sense to include only $SU(2)\times
U(1)$ invariant operators in one's list to start with, because the yet
unknown new physics lies at least a little above the electroweak
symmetry breaking scale. A host of such gauge invariant
higher-dimensional operators have been, and are being, analyzed with
considerable rigor, and now there exist limits on them, using data
ranging from electroweak precision measurements to global fits of LHC
results~\cite{Khachatryan:2014kca,GonzalezGarcia:1999fq,Han:2005pu,Desai:2011yj,Chen:2013kfa,Dekens:2013zca,Freitas:2012kw,McKeen:2012av,Godbole:2014cfa,
Godbole:2013saa,Biswal:2012mp,Low:2012rj,
Biswal:2009ar,Cakir:2013bxa,Godbole:2006eb,Godbole:2007cn,Biswal:2005fh,Christensen:2010pf,Barger:2003rs,
Han:2000mi,Plehn:2001nj,Banerjee:2013apa,Djouadi:2013qya,Anderson:2013afp,Ellis:2014dva,Chen:2014ona,Masso:2012eq,
Manohar:2006gz,Chang:2013cia,Gripaios:2013lea,Belusca-Maito:2014dpa,Corbett:2012dm,Corbett:2012ja,Falkowski:2014tna}.

Most of such analyses include higher-dimensional operators that
conserve charge conjugation (C), parity (P) and time-reversal
(T). However, there is evidence of C, P and CP-violation in weak
interactions~\cite{Agashe:2014kda}, and there are speculations about other sources of
CP-violation as well, especially with a view to explaining the baryon
asymmetry in our universe~\cite{Sakharov:1967dj,Riotto:1998bt}. The possibility of CP-nonconservation
cannot therefore be ruled out in the new physics currently sought
after.  Thus one may in principle also obtain higher-dimensional interaction
terms involving the Higgs and a pair of gauge bosons. The constraints
on such terms, and identification of regions in the parameter space where they 
can be phenomenologically significant, form the subject-matter of the present paper.

The CP-violating effective couplings, interestingly, are not
constrained by the oblique electroweak parameters at one-loop level up to ${\cal O}(\frac{1}{\Lambda ^2})$, 
where $\Lambda$ is the cutoff scale of the effective theory. The leading contributions to self-energy corrections to electroweak
gauge bosons at one-loop level occur at ${\cal O}(\frac{1}{\Lambda ^4})$. Therefore, electroweak precision (EWP) data are not expected 
to provide severe constraints on CP-odd parameters. 
In addition to this, Higgs-mediated event rates in various channels
receive contributions from these couplings at ${\cal O}(\frac{1}{\Lambda ^4})$. 
Thus they can also be constrained from
global fits of the LHC data. The strongest limits on them, however,
arise from the contributions to the electric dipole moments (EDMs) of the
neutron and the electron, both of which are severely restricted from
experiments. As we shall see in the following sections, a single
CP-violating operator taken at a time may in certain cases be limited to a very small
strength from the above constraints, while two or three such operators
considered together can have relatively larger, but highly correlated
coefficients. Some of these operators can have interesting phenomenological implications,
especially in the context of the LHC.

A study in similar lines can be found in Refs.\cite{Manohar:2006gz,Chang:2013cia,Gripaios:2013lea,Belusca-Maito:2014dpa,McKeen:2012av}. 
However, we have performed the most
comprehensive analysis, taking all the five possible dimension-6 CP-violating $VVh$ operators
($V = W,Z,\gamma$), which are not yet discussed thoroughly  in the literature. 
We provide the constraints obtained from the oblique electroweak parameters. 
The constraints coming from global fits of LHC data and electric dipole moments, for
two and three operators taken at a time, have been compared.
In addition to these we also provide the constraints on trilinear gauge 
boson couplings coming from LEP data on gauge boson pair production.

Our paper is organized as follows. In section 2, we provide details on the CP-violating gauge Higgs operators and 
derive Feynman rules for three point vertices of our interest. In section 3 we present the constraints on CP-violating
parameters coming from the precision electroweak measurements. Following this, we 
perform a global analysis of these parameters using LHC data on Higgs in section 4 and 
we discuss EDM constraints on CP-violating parameters in section 5. Section 6 
is devoted to discussion of results. Finally we summarize and conclude in section 7.

\section{Morphology of CP-violating Gauge-Higgs Operators}
In the effective Lagrangian approach that has been followed here, one can write a Lagrangian (${\cal L}_{\rm eff}$) comprising 
only of the SM fields, where the effects of the new physics that appear above the cutoff scale $\Lambda$ are encapsulated in higher 
dimensional gauge invariant operators. In general,
\begin{equation}
 {\cal L}_{\rm eff} = \sum_i \frac{f_i}{\Lambda^{d_i-4}} O^i,\label{eq:Leff}
\end{equation}
where $d_i > 4$ is the mass dimension of the operator $O^i$ and the dimensionless free parameter $f_i$ fixes the strength
of the corresponding operator. The operators constructed out of the Higgs doublet and the $SU(2)\times U(1)$ gauge fields 
are of even dimensions, and at the leading order they have mass dimension $d_i=6$.
The dimension six CP-even gauge invariant operators constructed out of the Higgs doublet ($\Phi$) and the electroweak gauge fields 
($B_\mu,\; W_\mu^a$), that modify the gauge-Higgs couplings are given as follows:
\begin{eqnarray}
 {O}_W &=& \frac{{f}_{W}}{\Lambda^2}(D_{\mu}\Phi)^\dagger\hat{{W}}^{\mu\nu}(D_{\nu}\Phi);~~
 {O}_B = \frac{{f}_{B}}{\Lambda^2}(D_{\mu}\Phi)^\dagger\hat{{B}}^{\mu\nu}(D_{\nu}\Phi); \nonumber \\ 
 {O}_{BB} &=& \frac{{f}_{BB}}{\Lambda^2}\Phi^\dagger\hat{B}^{\mu\nu}\hat{{B}}_{\mu\nu}\Phi;~~~~~~
 {O}_{WW} = \frac{{f}_{WW}}{\Lambda^2}\Phi^\dagger\hat{{W}}^{\mu\nu}\hat{W}_{\mu\nu}\Phi; \nonumber \\
 {O}_{BW} &=& \frac{{f}_{BW}}{\Lambda^2}\Phi^\dagger\hat{{B}}^{\mu\nu}\hat{W}_{\mu\nu}\Phi.  
 \end{eqnarray}\label{eq:O-cp-even}
In the above, we have defined $ \hat{B}_{\mu\nu} = i \frac{g'}{2} B_{\mu\nu}$ and  $ \hat{W}_{\mu\nu} = i \frac{g}{2} \tau^aW^a_{\mu\nu}$. 
$g$ and $g'$ are the electroweak coupling parameters corresponding to $SU(2)$ and $U(1)$ gauge groups respectively, and 
$\tau^a (a=1,2,3)$ are the three Pauli matrices.
% $g$ and $g'$ are the electroweak coupling parameters corresponding to $SU(2)$ and $U(1)$ gauge groups respectively.
We define the gauge covariant derivative as 
% \begin{equation}
$D_\mu~\equiv~\partial_\mu-i\frac{g}{2}\tau^aW_\mu^a-i\frac{g'}{2}YB_\mu$, 
where $Y$ is the hypercharge quantum number.
% \end{equation}
% where $\tau^a$ (a~=~1,2,3) are the three Pauli Matrices. 
With this choice of the definition of gauge covariant derivative, field strength tensor $W_{\mu\nu}^a$ is given by,
% 
% \begin{equation}
$W_{\mu\nu}^a~=~\partial_\mu W_\nu^a-\partial_\nu W_\mu^a + g\epsilon^{abc}W_\mu^bW_\nu^c$. 
% \end{equation}
% 
%  Out of these ${O}_{BW}$ is severely constrained by the precision electroweak measurements as it contributes to the 
%  $S$ parameter at tree level~\cite{GonzalezGarcia:1999fq}. 
The constraints on CP-even parameters and their collider implications have been studied extensively in the 
literature~\cite{GonzalezGarcia:1999fq,Han:2005pu,Masso:2012eq,Banerjee:2013apa,Corbett:2012dm,Corbett:2012ja,Falkowski:2014tna}. 
In this work we are interested in corresponding CP-violating dimension six 
gauge-Higgs operators. These are, 
\begin{eqnarray}
 \tilde{O}_W &=& \frac{\tilde{f}_{W}}{\Lambda^2}(D_{\mu}\Phi)^\dagger\hat{\tilde{W}}^{\mu\nu}(D_{\nu}\Phi);~~ 
 \tilde{O}_B = \frac{\tilde{f}_{B}}{\Lambda^2}(D_{\mu}\Phi)^\dagger\hat{\tilde{B}}^{\mu\nu}(D_{\nu}\Phi); \nonumber \\
 \tilde{O}_{BB} &=& \frac{\tilde{f}_{BB}}{\Lambda^2}\Phi^\dagger\hat{B}^{\mu\nu}\hat{\tilde{B}}_{\mu\nu}\Phi;~~~~~~ 
 \tilde{O}_{WW} = \frac{\tilde{f}_{WW}}{\Lambda^2}\Phi^\dagger\hat{\tilde{W}}^{\mu\nu}\hat{W}_{\mu\nu}\Phi \nonumber; \\
 \tilde{O}_{BW} &=& \frac{\tilde{f}_{BW}}{\Lambda^2}\Phi^\dagger\hat{\tilde{B}}^{\mu\nu}\hat{W}_{\mu\nu}\Phi,
 \end{eqnarray}\label{eq:O-cp-odd}
where, $\hat{\tilde{W}}^{\mu\nu}~=~\frac{1}{2}\epsilon^{\mu\nu\alpha\beta}\hat{W}_{\alpha\beta}$ 
and $\hat{\tilde{B}}^{\mu\nu}~=~\frac{1}{2}\epsilon^{\mu\nu\alpha\beta}\hat{B}_{\alpha\beta}$,
$\epsilon^{\mu\nu\alpha\beta}$ being the four-dimensional fully antisymmetric tensor with  $\epsilon^{0123}=1$.

In principle, the CP-even operators [Eq.(\ref{eq:O-cp-even})] could have been assumed to exist 
simultaneously with the CP-odd ones considered here. However, such an approach generates far too large 
a set of free parameters, where the signature of the CP-violating effective couplings would 
be drowned. Moreover, the CP-even operators are independent of the CP-odd ones (and vice versa); therefore, 
setting them to zero is a viable phenomenological approach.
We therefore postulate that the new physics above scale $\Lambda$ is such that {\it only 
CP-violating dimension six effective operators are appreciable}, and the corresponding CP-conserving 
ones are much smaller. Such a ``simplified approach,'' we reiterate, is unavoidable for unveiling 
CP-violating high scale physics, as has been recognized in the literature~\cite{Voloshin:2012tv,McKeen:2012av,Gavela:2014vra,Yepes:2015qwa}.
Studies focusing exclusively on the generation of CP-violating terms in specific new physics frameworks can also 
be found, an example being those in the context of extra space-time dimensions~\cite{Lim:1990bp,Lim:2009pj}. 

% For further discussions on this we refer the reader to section 6.

\begin{table}[t]
\begin{center}
\begin{tabular}{|c|c|}
  \hline
    & \\
  {\bf Coupling} & {\bf Effective coupling strength} \\
  & \\
  \hline
    & \\
  $C_{WWh}$    &    $(-\tilde{f}_{W}~-~2\tilde{f}_{WW})$\\
    & \\
  \hline 
   & \\
  $C_{ZZh}$ & $- 1/c^2_{W} \Big[c^2_{W}\tilde{f}_W+s^2_{W}\tilde{f}_B + 
  2(c^4_{W}\tilde{f}_{WW}~+~s^4_{W}\tilde{f}_{BB}) +2s^2_{W} c^2_{W} \tilde{f}_{BW}\Big]$ \\
   & \\
  \hline
    & \\
    $C_{\gamma\gamma h}$ & $-2s^2_{W}(\tilde{f}_{WW}+\tilde{f}_{BB}-\tilde{f}_{BW})$\\
      & \\
    \hline
      & \\
  $C_{\gamma Zh}$ & ${t_{W}}/{2}\Big[(-\tilde{f}_W + \tilde{f}_B)+4(s^2_{W}\tilde{f}_{BB}-c^2_{W}\tilde{f}_{WW})+
  2c_{2{W}}\tilde{f}_{BW}\Big]$\\
    & \\
      \hline 
        & \\
  $C_{WW\gamma}$ & $s_{W}/2(\tilde{f}_{W}+\tilde{f}_{B}+2\tilde{f}_{BW})$ \\ 
    & \\
      \hline
      & \\
       $C_{WWZ}$ & -$s_{W} t_{W}/2(\tilde{f}_{W}+\tilde{f}_{B}+2\tilde{f}_{BW})$ \\ 
    & \\
      \hline
        \end{tabular}
        \caption{CP-odd $VVh$ and $WWV$ coupling factors and their effective strengths. }
        \label{tab:Anomalous_vertices}
        \end{center}
 \end{table}

Since we focus on the extension of the SM through the inclusion of the CP-odd operators only, the full BSM 
Lagrangian looks like, 
\begin{equation}
{\cal L}_{\rm BSM} = {\cal L}_{\rm SM} + \tilde{O}_W +\tilde{O}_{WW} +\tilde{O}_B +\tilde{O}_{BB} +\tilde{O}_{BW},
\end{equation}\label{eq:L-bsm}
where ${\cal L}_{\rm SM}$ is the standard model Lagrangian. After the electroweak symmetry breaking, 
these CP-odd operators contribute to following three-point vertices of our interest\footnote{ The 
CP-odd operators considered here also contribute to four-point and five-point vertices like $VVhh$, 
$VVVh$ and $VVVhh$. However, as we will see in following sections, all the observables used in our 
analysis are sensitive to only three-point vertices at the leading order.},  
\begin{eqnarray}
{\cal L}_{WWh} &=& -\frac{g m_W}{\Lambda^2}
(\tilde{f}_{W} + 2\tilde{f}_{WW})  \epsilon^{\mu\nu\alpha\beta}k_{1\alpha}k_{2\beta} W^+_{\mu}(k_{1})W^-_{\nu}(k_{2})h(k)\label{eq:L-wwh}, \\ \nn \\
{\cal L}_{ZZh} &=&
- \frac{g m_W}{\Lambda^2}
 \Big[\frac{c^2_{W} \tilde{f}_W+s^2_{W} \tilde{f}_B}{c^2_{W}}+ 
\frac{2(c^4_{W}\tilde{f}_{WW}~+~s^4_{W}\tilde{f}_{BB})}{c^2_{W}} + 2s^2_{W}\tilde{f}_{BW}\Big] \nonumber \\
&& \epsilon^{\mu\nu\alpha\beta}k_{1\alpha}k_{2\beta} Z_{\mu}(k_{1})Z_{\nu}(k_{2})h(k)\label{eq:L-zzh},  \\ \nn \\
{\cal L}_{\gamma \gamma h} &=&-2\Big(\frac{g m_W}{\Lambda^2}\Big)s^2_{W}(\tilde{f}_{WW}+\tilde{f}_{BB}-\tilde{f}_{BW})\nonumber\\
&&\epsilon^{\mu\nu\alpha\beta}k_{1\alpha}k_{2\beta}A_{\mu}(k_1)A_{\nu}(k_2)h(k)\label{eq:L-aah}, 
%  \\ \nn \\
 \end{eqnarray}
 \begin{eqnarray}
  {\cal L}_{\gamma Zh}&=&\Big(\frac{g m_W}{ 2\Lambda^2}\Big)t_{W} 
 \Big[(-\tilde{f}_W +\tilde{f}_B)+4(s^2_{W}\tilde{f}_{BB}-c^2_{W}\tilde{f}_{WW})+2c_{2{W}}\tilde{f}_{BW}\Big] \nonumber\\
&& \epsilon^{\mu\nu\alpha\beta}k_{1\alpha}k_{2\beta}A_{\mu}(k_1)Z_{\nu}(k_2)h(k)\label{eq:L-azh},
 \\ \nonumber\\
{\cal L}_{WW\gamma} &=&\Big(\frac{g {m_W^2}}{2\Lambda^2}\Big)s_{W}(\tilde{f}_{W}+\tilde{f}_{B}+2\tilde{f}_{BW})\nonumber\\
 && \epsilon^{\mu\nu\alpha\beta}k_{\beta}W^+_{\mu}(k_1)W^-_{\nu}(k_2) A_{\alpha}(k)\label{eq:L-wwa},  \\ \nn \\
{\cal L}_{WWZ} &=&-\Big(\frac{g {m_W^2}}{2\Lambda^2}\Big)(s_{W} t_{W})(\tilde{f}_{W}+\tilde{f}_{B}+2\tilde{f}_{BW})\nonumber\\
 && \epsilon^{\mu\nu\alpha\beta}k_{\beta}W^+_{\mu}(k_1)W^-_{\nu}(k_2) Z_{\alpha}(k)\label{eq:L-wwz}. 
\end{eqnarray}
In the above equations $s_{W} ={\rm sin }{\theta_W}$, $c_{W} ={\rm cos}{\theta_W}$, 
$t_{W} ={\rm tan}{\theta_W}$, and $c_{2W} = {\rm cos}{2\theta_W}$, where $\theta_W$ is the Weinberg angle.
Here $k$s are the four-momenta of the fields that enter the vertex. We have taken all momenta to be inflowing toward the three-point vertex 
in establishing the Feynman rules. 
From the list of CP-odd interaction vertices shown above, one can observe a general tensor structure of the form 
$\epsilon^{\mu\nu\alpha\beta}k_{1\alpha}k_{2\beta}$ 
in $VVh$ vertices and a general tensor structure of the form 
$\epsilon^{\mu\nu\alpha\beta}k_{\beta}$ 
in trilinear gauge boson couplings ($WWV$). Because of this
the CP-odd couplings are linear combinations of the parameters $\tilde f_i$.
Note that we have not included the CP-odd operator involving gluon-Higgs coupling, 
\begin{equation}
 \tilde O_{GG} = \frac{\tilde{f}_{GG}}{\Lambda^2}\Phi^\dagger\hat{\tilde{G}}^{\mu\nu}\hat{G}_{\mu\nu}\Phi.\label{eq:O-ggh} 
\end{equation}
This operator introduces a $\theta_{QCD}$ term~\cite{Peccei:1977hh,Peccei:1977ur}, and it is severely constrained by the experimental measurement of neutron 
EDM~\cite{Agashe:2014kda}. 
In Table~\ref{tab:Anomalous_vertices}, we list various couplings and their effective strengths ignoring the overall dimension full factor of
$\frac{g m_W}{\Lambda^2}$ in $C_{VVh}$ and the dimensionless factor of $\frac{g m_W^2}{\Lambda^2}$ in $C_{WWV}$ couplings. Note that only 
$ZZh$ and $\gamma Zh$ couplings receive contribution from all five CP-odd operators. The operators which contribute to 
$WW\gamma$ also contribute to
$WWZ$ and these couplings are related by, $C_{WWZ} = -t_W C_{WW\gamma}$.

 \section{Constraints from Electroweak Precision (EWP) Data}

\begin{figure}[t]
  \begin{center}
\includegraphics[width = 0.8\linewidth]{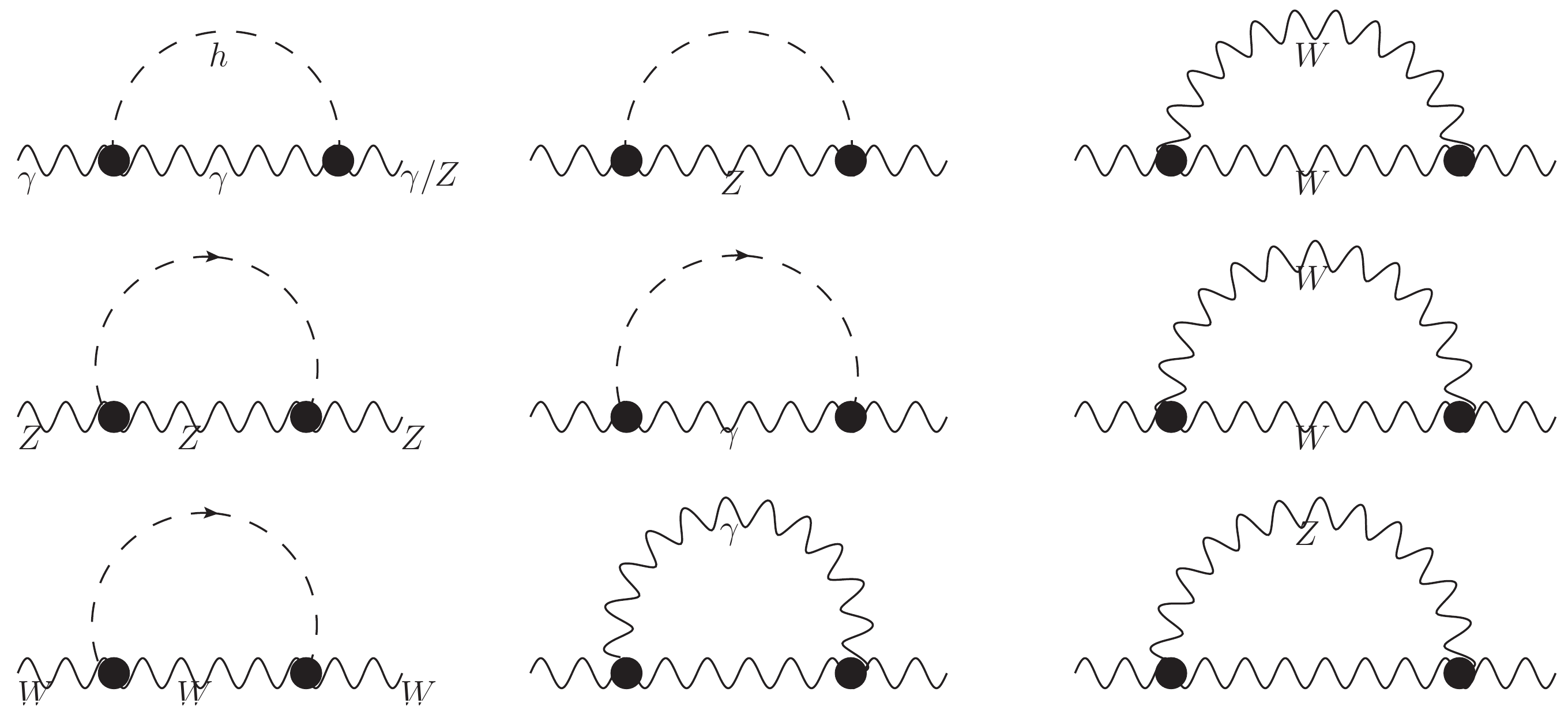}
\caption{One-loop self energy corrections to electroweak vector bosons (oblique corrections) in presence of CP-odd operators. The blobs show 
the effective CP-odd vertices. These corrections are of ${\cal O}(1/\Lambda^4)$.}
\label{fig:SelfEnegy-OneLoop}
  \end{center}
\end{figure}
 
We note that unlike some of the CP-even (D=6) operators, the CP-odd operators do not contribute to 
the gauge boson propagator 
corrections at tree level, hence are not expected to receive severe bounds from the electroweak precision
data. This is due to the antisymmetry of the epsilon tensor which is present 
in all CP-odd operators. In fact, because of the same reason, all quantum corrections to gauge boson two-point 
functions up to ${\cal O}(\frac{1}{\Lambda^2})$ vanish~\footnote{These corrections are proportional to 
$\epsilon^{\mu\nu\alpha\beta}p_{\alpha}p_{\beta}$ ($p$ being the four-momentum of gauge boson) which is zero.} 
and first nonzero contributions due to CP-odd 
operators appear at ${\cal O}(\frac{1}{\Lambda^4})$. As we will see in section 4, the CP-odd couplings 
contribute to observables related to LHC Higgs data at this order. It would be interesting to discuss 
the implications of the electroweak precision 
measurement constraints on the parameters of CP-odd operators.
% that to what extent the electroweak precision measurements can constrain the parameters.

 \begin{figure}
%  \begin{center}
 \includegraphics[width = 0.30\linewidth]{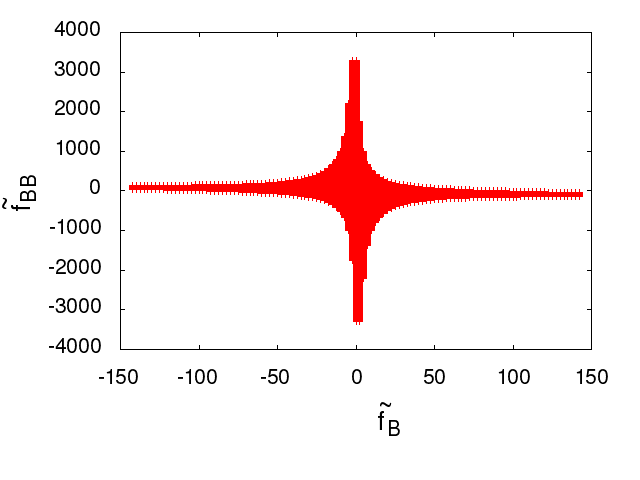}
 \includegraphics[width = 0.30\linewidth]{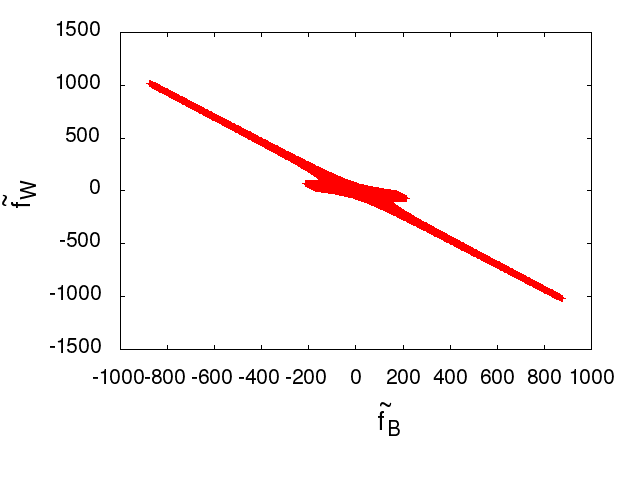}
 \includegraphics[width = 0.30\linewidth]{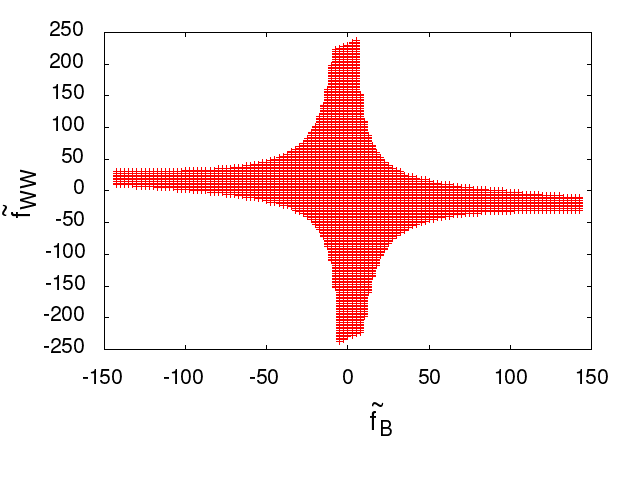}
 \includegraphics[width = 0.30\linewidth]{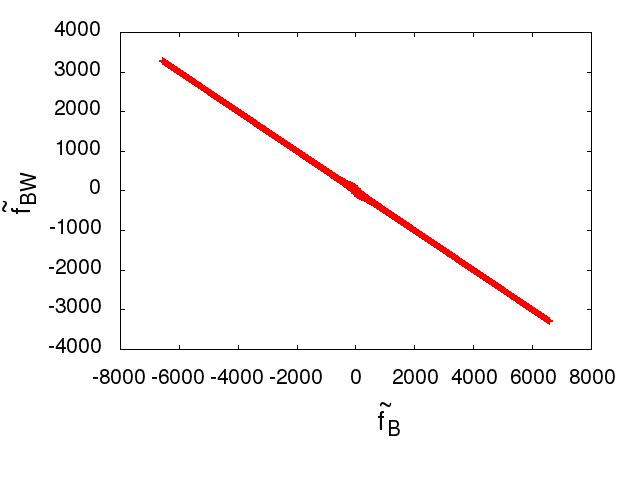}
 \includegraphics[width = 0.30\linewidth]{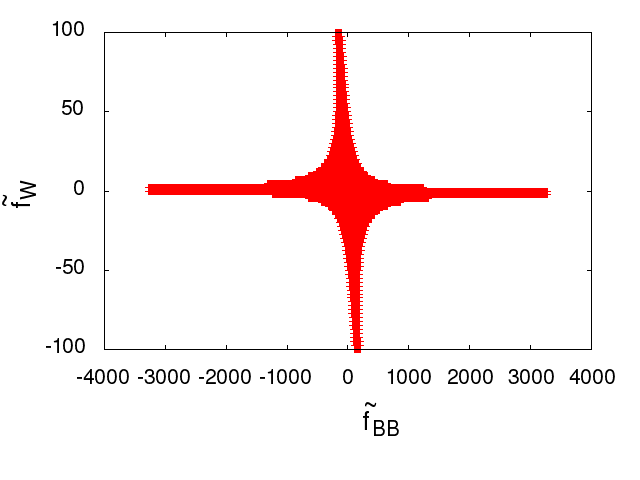}
 \includegraphics[width = 0.30\linewidth]{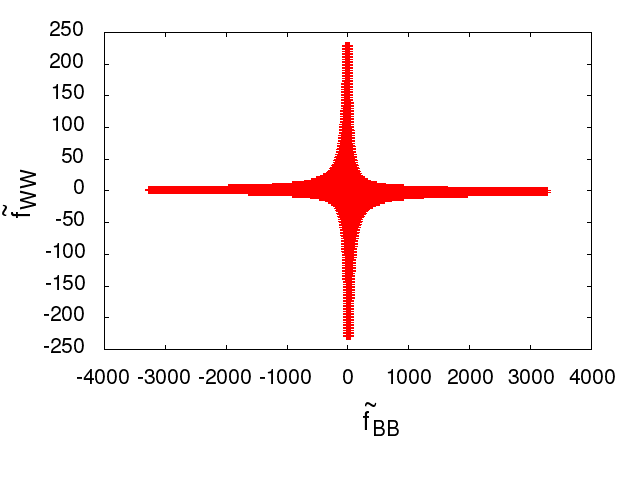}
 \includegraphics[width = 0.30\linewidth]{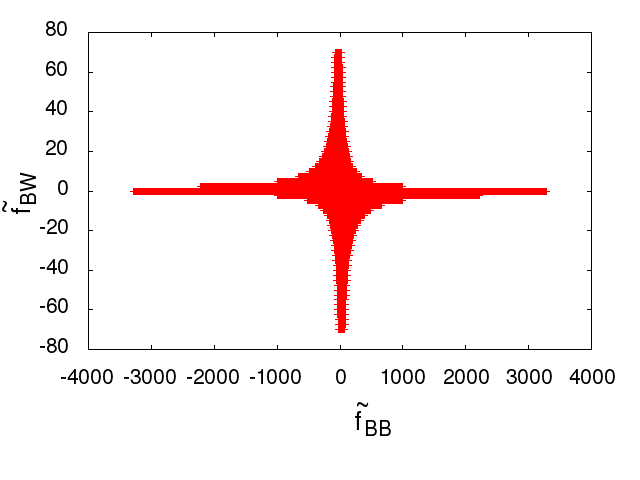}
 \includegraphics[width = 0.30\linewidth]{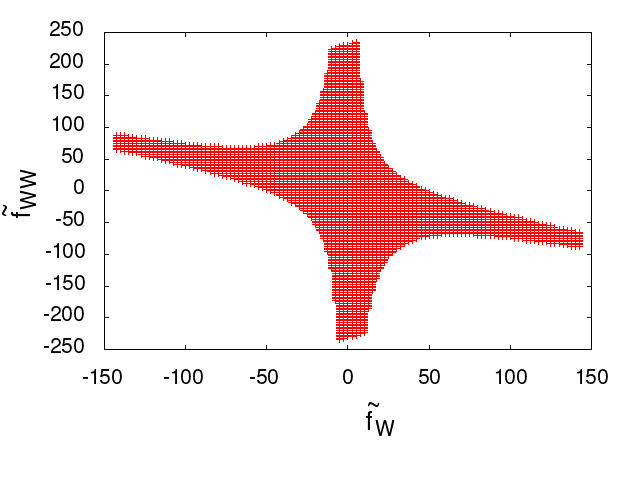}
 \includegraphics[width = 0.30\linewidth]{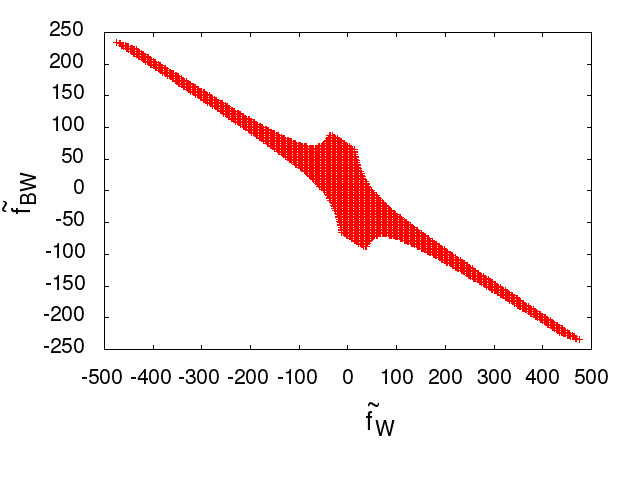}
 \includegraphics[width = 0.30\linewidth]{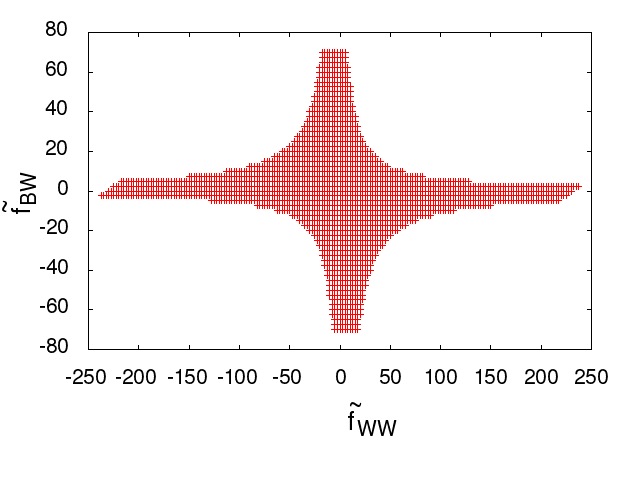}
 \caption{Constraints from electroweak precision data keeping two parameters 
nonzero at a time and for $\Lambda = 1 $ TeV.}
%  \end{center}
 \label{fig:STU-plots}
  \end{figure}

It is well known that the dominant effects of new physics can be conveniently parametrized in terms 
of Peskin-Takeuchi parameters~\cite{Peskin:1991sw}. These are related to the gauge boson two-point functions as,  
\begin{eqnarray}
 \alpha S &=& 4 c_W^2 s_W^2 \Big(  \Pi_{ZZ}'(0) - \Pi_{\gamma\gamma}'(0) 
 -\frac{c_W^2 -s_W^2}{s_Wc_W} \Pi_{\gamma Z}'(0) \Big)\label{eq:def-S} \\
 \alpha T &=&   \frac{\Pi_{WW}(0)}{m_W^2} - \frac{\Pi_{ZZ}(0)}{m_Z^2}\label{eq:def-T} \\
 \alpha U &=& 4 s_W^2 \Big( \Pi_{WW}'(0) -c_W \Pi_{ZZ}'(0) 
 -s_W^2\Pi_{\gamma\gamma}'(0) -2c_Ws_W\Pi_{\gamma Z}'(0) \Big)\label{eq:def-U}
\end{eqnarray}
where, $\Pi_{V_1V_2}(p^2)$ and $\Pi_{V_1V_2}'(p^2)$ are the $g^{\mu\nu}$ part of the two-point function and 
 its derivative with respect to $p^2$, respectively. The relevant one-loop Feynman diagrams are shown in Fig.~\ref{fig:SelfEnegy-OneLoop}. 
 We have regularized ultraviolet (UV) singularities of these diagrams in dimensional regularization (DR). 
 The expressions for $\Pi_{V_1V_2}(p^2)$ in terms of standard one-loop scalar functions are given in Appendix~\ref{App:AppendixA}. 
 We find that $\Pi_{\gamma\gamma}(0) = \Pi_{\gamma Z}(0) = 0 $ which is expected due to the transverse nature of the photon.
 The renormalization of UV singularity is carried out in $\overline{\rm MS}$ scheme which introduces scale dependence in 
 these expressions.
 We have identified the renormalization scale  with the cutoff scale $\Lambda$. Thus the gauge boson two-point 
 functions also have ${\rm ln}(\Lambda)$ dependence apart from the overall $1/\Lambda^4$ dependence coming from CP-odd
 couplings. Because of this 
 an explicit choice of $\Lambda$ is necessary in deriving the EWP constraints on $\tilde f_i$s.

For $\Lambda = 1 $ TeV, the Peskin-Takeuchi parameters due to CP-odd couplings 
are given by, 
\begin{eqnarray}
  S &=& (- 3.36\; C_{\gamma\gamma h}^2 - 1.28\; C_{\gamma Zh}^2 + 4.64\; C_{ZZh}^2 
                    - 4.49\; C_{\gamma\gamma h} C_{\gamma Zh} - 6.21\; C_{\gamma Zh} C_{ZZh} ) \times 10^{-5}\label{eq:S-cp-odd}  \\
  T &=& -9.74 \times 10^{-5}\; C_{WW\gamma}^2\label{eq:T-cp-odd} \\
  U &=& (-0.960\;C_{\gamma\gamma h}^2 - 4.69\;C_{\gamma Zh}^2  - 4.64\;C_{ZZh}^2 + 5.67\;C_{WWh}^2 + 2.76\;C_{WW\gamma}^2 \nn \\
             && -\; 3.59\;C_{\gamma\gamma h} C_{\gamma Zh} - 4.96\;C_{\gamma Zh} C_{ZZh} )  \times 10^{-5}\label{eq:U-cp-odd}.
\end{eqnarray}
We can also express them in terms of $\tilde f_i$s using their relation with $C_i$s given in Table~\ref{tab:Anomalous_vertices}.
The experimental limits on $S,T$ and $U$ parameters are obtained by fitting the data on various electroweak observables with 
these parameters. The limits are~\cite{Agashe:2014kda}, 
\begin{eqnarray}
  S = -0.03 \pm 0.10, \:
  T = 0.01 \pm 0.12, \:
  U = 0.05 \pm 0.10.
\end{eqnarray}
In Eqs. (~\ref{eq:S-cp-odd}), (\ref{eq:T-cp-odd}) and (\ref{eq:U-cp-odd}) the coefficients of various couplings 
are $\sim 10^{-5}$ suggesting that the EWP 
constraints cannot be very strong. Therefore, we only consider the case where any
two out of five parameters are nonzero. 
In Fig.~\ref{fig:STU-plots}, we display the allowed range for CP-odd parameters which 
satisfy the above limits on $S, T$ and $U$ parameters for all ten sets of two parameters 
taken together. Here we have varied parameters 
freely to ensure that we obtain a bounded region. We note that large values of ${\cal O}(1000)$ for 
$\tilde f_{BB}$ are always allowed. On the other hand the allowed range for $\tilde f_{WW}$ 
never goes beyond 250. Allowed values for all other parameters can be of ${\cal O}(100-1000)$. 
Some of these observations can be understood once we express the $S, T$ and $U$ parameters in 
terms of $\tilde f_i$s. As we turn on other parameters, these constraints become weaker. 
Also, for a larger cutoff scale the allowed parameter space grows as one would expect.

\section{Constraints from LHC data}
The presence of CP-odd operators introduces modifications in the strength of the gauge-Higgs couplings, and hence changes the 
Higgs production and decay rates in channels involving these couplings. Since we are interested in CP-even 
observables, the SM $VVh$ couplings which are CP-even, do not interfere with the CP-odd $VVh$ couplings.
Hence the lowest order (tree level) modifications to the decay widths ($\Gamma$) and
production cross sections ($\sigma $) are of the order $\frac{1}{\Lambda^4}$. To quantify these changes we define the following ratios
for various decay and production channels,
\begin{eqnarray}
 \alpha_{Y}  &=& \frac{\Gamma^{\rm BSM}(h\rightarrow Y)}{\Gamma^{\rm SM}(h\rightarrow Y)}\\
  \gamma_{X} &=& \frac{\sigma^{\rm BSM}(X\rightarrow h)}{\sigma^{\rm SM}(X\rightarrow h)}
\end{eqnarray}
where $Y$ and $X$ are used to label the final state and initial state particles in the Higgs decay and production channels respectively. 
% The greek letters $\Gamma$ and $\sigma$ denote the decay width and the cross section.

\subsection{Higgs decay channels}
In the SM, the 126 GeV Higgs boson predominantly decays into $b{\bar b}$ and $WW^*$ followed by $gg,~\tau^+\tau^-~,c{\bar c}$ and 
$ZZ^*$. It also decays to $\gamma\gamma$, $\gamma Z$ and $\mu^+\mu^-$ with much suppressed rates. Out of these, $h \to gg$, $h \to \gamma\gamma$ 
and $h \to \gamma Z$ are loop-induced decay modes and hence are sensitive to new physics. 
The decay channels which are affected by the CP-odd operators are $h\rightarrow \gamma\gamma$, $h\rightarrow \gamma Z$,
$h\rightarrow WW^*$ and  $h\rightarrow ZZ^*$. 
The expressions for the ratio of the decay widths, $\alpha_{ij}$ in various two body Higgs decay channels are as follows
\footnote{ We disagree with the CP-odd part of the analytic expression for $\alpha_{\gamma Z}$ in Eq. (3.17) of Ref.~\cite{Chang:2013cia}. 
The correct expression for CP-odd term in the notations of Ref.~\cite{Chang:2013cia} turns out to be, 
\begin{equation}
 \alpha_{\gamma Z} =  1 + \Big| \frac{4\sqrt{2}\pi^2{\tilde a_2}}{G_F\Lambda^2 s_{\rm W}^2(A_F+A_W)} \Big|^2,
\end{equation}
where ${\tilde a_2}$ can be identified with the factor $\frac{1}{2}C_{\gamma Z h}$ in our notation.}:
\begin{eqnarray}
\alpha_{\gamma\gamma} &=& 1 + 2.84\Big(\frac{C^2_{\gamma\gamma h}}{\Lambda^4}\Big)\label{eq:aa}\\
\nonumber\\
\alpha_{\gamma Z} &=&  1 + 0.856\Big(\frac{C^2_{\gamma Zh}}{\Lambda^4}\Big)\label{eq:az} 
% \\ \nonumber\\
% 
\end{eqnarray}
 \begin{eqnarray}
\alpha_{WW^*} &=&  1 + 3.35\times10^{-6}\Big(\frac{C^2_{WW h}}{\Lambda^4}\Big)\\
\nonumber\\
\alpha_{2\ell2\nu} &=&  1 + 3.56 \times10^{-6}\Big(\frac{C^2_{WW h}}{\Lambda^4}\Big)\label{eq:2l2nu} 
 \\ \nonumber\\
\alpha_{ZZ^*} &=& 1 + 1.40\times10^{-6}\Big(\frac{C^2_{ZZ h}}{\Lambda^4}\Big)\\
\nonumber\\
\alpha_{4\ell} &=& 1 + 1.54\times10^{-6}\Big(\frac{C^2_{ZZ h}}{\Lambda^4}\Big)\label{eq:4l}
% \nonumber
% \frac{\Gamma^{\rm BSM}}{\Gamma^{\rm SM}}(h \to WW^*\to 2l2\nu) &=& 1 + 3.51({\cred 3.56})\times10^{-6}\Big(\frac{C^2_{WW h}}{\Lambda^4}\Big)\\
% \nonumber\\
% \frac{\Gamma^{\rm BSM}}{\Gamma^{\rm SM}}(h \to ZZ^*\to l^+ l^- l^{'+} l^{'-}) &=& 1 + 1.66({\cred 1.54})\times10^{-6}\Big(\frac{C^2_{ZZ h}}{\Lambda^4}\Big)\\
\end{eqnarray}
Since the set of gauge-Higgs operators considered in our analysis do not 
 alter the Higgs coupling with gluons ($g$) and fermions ($f$), we have $\alpha_{f f} = \alpha_{gg}$ = 1. The ratios $\alpha_{2\ell2\nu}$ [Eq.(~\ref{eq:2l2nu})]
 and $\alpha_{4\ell}$ [Eq.(~\ref{eq:4l})] correspond to the $ h \to WW^* \to 2\ell2\nu$ and $ h \to ZZ^* \to 4\ell$ respectively. 
 Here $\ell$ stands for electron and muon, and $\nu$ for corresponding neutrinos. The ratios $\alpha_{WW^*}$ and $\alpha_{ZZ^*}$ which 
 include both leptonic and hadronic decays of $W$ and $Z$ bosons are 
 used in calculating modified total Higgs decay width. 
 As mentioned earlier, the modifications to Higgs partial decay widths at leading order are ${\cal O}(1/\Lambda^4)$. It is in contrast to the case of 
 CP-even dimension six gauge-Higgs operators where such modifications occur at ${\cal O}(1/\Lambda^2)$. Unlike $WWh$ and 
 $ZZh$ couplings, the $\gamma\gamma h$ and $\gamma Z h$ couplings are loop-induced in the SM. In the presence of CP-odd 
 operators these vertices receive contributions at tree level. This explains the relatively large coefficients 
 in the expressions for $\alpha_{\gamma\gamma}$ [Eq.(~\ref{eq:aa})] and $\alpha_{\gamma Z}$ [Eq.(~\ref{eq:az})] as compared to the 
 other decay width ratios. This would imply most stringent constraints on the parameters contributing to these 
 decay channels.
  For further discussion on CP-odd vs. CP-even operators, we refer the reader to section 6.

\subsection{Higgs production channels}
At the LHC, the dominant mode to produce Higgs boson is  gluon-gluon fusion (GGF) mediated 
by a top quark loop. The 
other major production channels include: vector boson fusion (VBF), associated production with 
a weak boson ($Vh$) and associated production with a pair of top quark ($t{\bar t} h$).
Except GGF and $t{\bar t}h$ production channels, all other channels are affected in presence of 
anomalous gauge-Higgs CP-odd vertices.
Like the decay width ratios, the production cross section ratios also receive modifications at ${\cal O}(1/\Lambda^4)$. 
The ratios of the Higgs production cross sections, $\gamma_X$ in various channels at $\sqrt{S}=$ 8(7) TeV LHC are given below. 

\begin{eqnarray}
\gamma_{p p \to Wh} &=& 1 + 5.61(5.37) \times 10^{-4}\Big(\frac{C^2_{WW h}}{\Lambda^4}\Big)\\
\nonumber\\
\gamma_{p p \to Wh \to hl\nu} &=& 1 + 5.67( 5.16) \times 10^{-4}\Big(\frac{C^2_{WW h}}{\Lambda^4}\Big)\\
\nonumber\\
\gamma_{p p \to Zh} &=& 1 + 4.09(3.92)\times 10^{-4}\Big(\frac{C^2_{ZZh}}{\Lambda^4}\Big)
                       +  2.45(2.32)\times10^{-4}\Big(\frac{C^2_{\gamma Zh}}{\Lambda^4}\Big) \nonumber\\
                     &&+~    2.55(2.44)\times 10^{-4}\Big(\frac{C_{ZZh}}{\Lambda^2}\Big)\Big(\frac{C_{\gamma Zh}}{\Lambda^2}\Big)\label{eq:Zh}
%                      \\ \nonumber\\
%  \gamma_{p p \to Zh \to h2l} &=& 
%  \nonumber\\
\end{eqnarray}
\begin{eqnarray}
\gamma_{\rm VBF} &=& 1 + 7.02(5.62)\times 10^{-6}\Big(\frac{{\tilde f}_{B}^2}{\Lambda^4}\Big) 
   + 1.50(1.44)\times10^{-4}   \Big(\frac{{{\tilde f}_W}^2}{\Lambda^4}\Big)
   + 1.84(1.80)\times10^{-5}   \Big(\frac{{\tilde f}_{BB}^2}{\Lambda^4}\Big)  \nonumber\\
  &&+~ 6.98(6.75)\times10^{-4} \Big(\frac{{\tilde f}_{WW}^2}{\Lambda^4}\Big) 
   + 4.39(4.38) \times 10^{-5} \Big(\frac{{\tilde f}_{BW}^2}{\Lambda^4}\Big) 
    - 1.32(1.14)\times10^{-5} \Big(\frac{{\tilde f}_B {\tilde f}_W}{\Lambda^4}\Big) \nonumber\\
  && +~ 8.96(22.2) \times 10^{-7} \Big(\frac{{\tilde f}_B {\tilde f}_{BB}}{\Lambda^4}\Big) 
   - 5.06(5.03) \times 10^{-5} \Big(\frac{{\tilde f}_B {\tilde f}_{WW}}{\Lambda^4}\Big) \nonumber\\
  && +~ 2.63(2.59) \times 10^{-5} \Big(\frac{{\tilde f}_B {\tilde f}_{BW}}{\Lambda^4}\Big) 
   +~ 8.98(9.21)\times 10^{-7}\Big(\frac{{\tilde f}_W {\tilde f}_{BB}}{\Lambda^4}\Big) \nonumber\\
   &&+~ 6.22(5.98)\times 10^{-4}\Big(\frac{{\tilde f}_W {\tilde f}_{WW}}{\Lambda^4}\Big) 
   - 2.46(2.71) \times 10^{-5} \Big(\frac{{\tilde f}_W {\tilde f}_{BW}}{\Lambda^4}\Big) \nonumber\\
     &&-~ 3.51(3.60) \times 10^{-5}  \Big(\frac{{\tilde f}_{BB} {\tilde f}_{BW}}{\Lambda^4}\Big) 
   + 3.27(3.85) \times 10^{-5} \Big(\frac{{\tilde f}_{BB} {\tilde f}_{WW}}{\Lambda^4}\Big) \nonumber\\
   &&-~1.43(1.45) \times 10^{-4}  \Big(\frac{{\tilde f}_{WW} {\tilde f}_{BW}}{\Lambda^4}\Big) \label{eq:vbf} \\
\nonumber\\
\gamma_{pp \to tth} &=& \gamma_{\rm GGF} = 1.
\end{eqnarray}
% {\cred put $\Lambda$ in 34.}
% 
These expressions have been obtained by computing the SM and BSM cross sections 
at tree level using {\tt Madgraph}~\cite{Alwall:2014hca} under the assumption that the K-factors (due to higher order corrections) are same in the  
SM and BSM cases. For that we have implemented our effective Lagrangian in {\tt FeynRules}~\cite{Christensen:2008py} and used the generated 
UFO model file in {\tt Madgraph}. The cross sections have been calculated using {\tt cteq6l1} parton distribution functions~\cite{Pumplin:2002vw} 
and with default settings for renormalization and factorization scales.

We would like to point out that there is an additional diagram which contributes to $pp \to Zh$ due to tree level CP-odd $\gamma Zh$ coupling. Similarly, in VBF 
channel additional diagrams appear due to both $\gamma\gamma h$ and $\gamma Zh$ couplings. Because of a different parametrization, this information 
is not explicit in the expression for VBF. We find this parametrization more convenient in terms of evaluating the coefficients in Eq.(~\ref{eq:vbf}). 
Also, the VBF coefficients reported above do not have any $Vh$ contamination and this can be ensured in {\tt Madgraph} at the process 
generation level. 
One can notice that the modifications induced by the CP-odd operators are
relatively weak because the SM cross sections are already tree level effects in the modified production channels.

One important fact in relation to these production cross section ratios ($\gamma$s) is that the numerical coefficients present
in these expressions are very much cut dependent. This is associated with the fact that the anomalous couplings induced by the gauge-Higgs
operators have a different Lorentz structure (therefore, different kinematic dependence) than their SM counterparts. 
Hence in general the SM and BSM cut efficiencies are not the same for a given process. The differences
between the two become more pronounced for higher values of the CP-odd couplings. But for reasonably low values of the same 
one can still work under the approximation that the two cut efficiencies are the same.
In this work we have taken this approximation into consideration, and taken only default cuts in {\tt Madgraph} to simulate
any production or decay channel. Since we have taken data only from individual production channels 
and not from combined channel data (e.g. $Vh$ combined channel) in our chi-square analysis, this approximation finds a stronger footing.

\begin{table}[t]
 \begin{center}
\begin{tabular}{|c|c|c|}
  \hline
   &  & \\
  {\bf Production channel} & {\bf 8 TeV cross section (pb)} & {\bf 7 TeV cross section (pb)}\\
   & &  \\
  \hline
    &  &\\
  $GGF$   &18.97   &14.89 \\
    &  &\\
  \hline 
   &  &\\
    $VBF$   &1.568   &1.211 \\
    &  &\\
  \hline 
   &  &\\
    $Wh$   &0.686   & 0.563\\
    &  &\\
  \hline 
   &  &\\
    $Zh$   &0.405   &0.327 \\
    &  &\\
  \hline 
   &  &\\
    $t{\bar t}h$   &0.1262  &0.0843 \\
    &  &\\
  \hline 
   &  &\\
    $b{\bar b}h$   &0.198   &0.152 \\
    &  &\\
   \hline
   
   \end{tabular}
  \end{center}
  \caption{Higgs production cross section in the SM~\cite{twiki_CERN}.}\label{tab:smXS}
  \end{table}

 \subsection{Global analysis }
 
The quantitative measure of the difference between the Higgs data from the LHC, and its corresponding SM predictions is
given by what we call the signal strength, defined as,
\begin{equation}
\mu^{X,Y} =  \frac{\sigma^{\rm BSM}(X\rightarrow h){\rm BR}^{\rm BSM}(h\rightarrow Y)}{\sigma^{\rm SM}(X\rightarrow h){\rm BR}^{\rm SM}(h\rightarrow Y)}\label{eq:signal-strength}
\end{equation}
where, ${\rm BR}(h\rightarrow Y) = \frac{\Gamma(h\rightarrow Y)}{\Gamma_{total}} $ is the branching ratio for Higgs decaying into $Y$ final 
state,
and ${\Gamma_{total}}$ is the total Higgs decay width. For 126 GeV Higgs boson $\Gamma^{\rm SM}_{total} =4.2$ MeV.
The total Higgs decay width in the BSM construct $\Gamma^{\rm BSM}_{total}$ 
can be expressed in terms of the SM total Higgs decay width $\Gamma^{\rm SM}_{total}$ by,
\begin{equation}
 \Gamma^{\rm BSM}_{total} = S_{total} \Gamma^{\rm SM}_{total}.
\end{equation}
$S_{total}$ is given in terms of the various branching fractions of the Higgs in the SM as,
\begin{eqnarray}
 S_{total} &\sim& {\rm BR}^{\rm SM}_{bb} + {\rm BR}^{\rm SM}_{cc} + {\rm BR}^{\rm SM}_{\tau\tau} + \alpha_{\gamma\gamma}{\rm BR}^{\rm SM}_{\gamma\gamma} + 
 \alpha_{\gamma Z}{\rm BR}^{\rm SM}_{\gamma Z} \nn \\
 &&+~ \alpha_{WW^*}{\rm BR}^{\rm SM}_{WW^*} + \alpha_{ZZ^*}{\rm BR}^{\rm SM}_{ZZ^*} + \alpha_{gg}{\rm BR}^{\rm SM}_{gg}
\end{eqnarray}
which becomes on solving,
\begin{equation}
S_{total}\sim 0.736 + 0.0023\alpha_{\gamma\gamma} + 0.0016\alpha_{\gamma Z} + 0.23\alpha_{WW^*} + 0.029 \alpha_{ZZ^*}.\label{eq:S-total}
\end{equation}
The SM branching fractions for 126 GeV Higgs are taken from \cite{twiki_CERN}.
The signal strength in Eq.(~\ref{eq:signal-strength}) can be rewritten in a compact form using the 
decay and the productions cross section ratios defined above, 
\begin{eqnarray}
 \mu^{X,Y} = \gamma_X \frac{\alpha_Y}{S_{total}}.
\end{eqnarray}

\begin{table}
 \begin{center}
\begin{tabular}{|c|c|c|c|}
  \hline
%    &  & &\\
  {\bf Production channel} & {\bf Decay channel} & {\bf Signal strength}& {\bf Energy in TeV }\\
  &  & & ({\bf Luminosity in $fb^{-1}$})\\
%   & & & \\
  \hline
    & & &\\
  ${\rm GGF~(ATLAS})$   &  $h\to \gamma\gamma$ & $1.32 \pm 0.38$ ~ \cite{Aad:2014eha} & 7(4.5) + 8(20.3)\\
    & & &\\
  \hline 
  & & &\\
  ${\rm VBF~(ATLAS})$  &  $h\to \gamma\gamma$ & $0.8 \pm 0.7$ ~ \cite{Aad:2014eha}& 7(4.5) + 8(20.3)\\
    & & &\\
     \hline
      & & &\\ 
  $W h~({\rm ATLAS})$  &  $h\to \gamma\gamma$ & $1.0 \pm 1.6$ ~ \cite{Aad:2014eha}& 7(4.5) + 8(20.3) \\
    & & &\\
  \hline  
   & & &\\
  $Z h~({\rm ATLAS})$  &  $h\to \gamma\gamma$ & $0.1^{+3.7}_{-0.1}$ ~ \cite{Aad:2014eha}& 7(4.5) + 8(20.3) \\
    & & &\\
  \hline 
    & & &\\
  $t{\bar t}h~({\rm ATLAS})$  &  $h\to \gamma\gamma$ & $1.6^{+2.7}_{-1.8}$ ~ \cite{Aad:2014eha}& 7(4.5) + 8(20.3) \\
    & & &\\
  \hline 
   & & &\\
  ${\rm GGF ~(CMS)}$  &  $h\to \gamma\gamma$ & $1.12^{+0.37}_{-0.32}$ ~ \cite{Khachatryan:2014ira}& 7(5.1) + 8(19.7) \\
    & & &\\
    \hline    
   & & &\\
  ${\rm VBF ~(CMS)}$  &  $h\to \gamma\gamma$ & $1.58^{+0.77}_{-0.68}$ ~ \cite{Khachatryan:2014ira}& 7(5.1) + 8(19.7) \\
    & & &\\
  \hline 
 & & &\\
  $t{\bar t}h ~({\rm CMS})$  &  $h\to \gamma\gamma$ & $2.69^{+2.51}_{-1.81}$ ~ \cite{Khachatryan:2014ira}& 7(5.1) + 8(19.7) \\
    & & &\\
  \hline 
  & & &\\
${\rm GGF}+t{\bar t}h+b{\bar b}h ~ ({\rm ATLAS})$  &  $h\to ZZ^* \to 4\ell$ & $1.7^{+0.5}_{-0.4}$ ~ \cite{Aad:2014eva}& 7(4.5) + 8(20.3) \\
    & & & \\
  \hline 
  & & &\\
  ${\rm GGF}+t{\bar t}h~ ({\rm CMS})$  &  $h\to ZZ^* \to 4\ell$ & $0.8^{+0.46}_{-0.36}$ ~ \cite{Chatrchyan:2013mxa}& 7(5.1) + 8 (19.7) \\
  & & &\\
  \hline
  & & &\\
  ${\rm GGF ~(ATLAS)}$  &  $h\to WW^* \to 2\ell2\nu$ & $1.01^{+0.27}_{-0.25}$ ~ \cite{ATLAS:WW*_2l2v} & 7(4.5) + 8(20.3)\\
    & & & \\
  \hline 
  & & &\\
  $ {\rm VBF ~(ATLAS)}$  &  $h\to WW^* \to 2\ell2\nu$ & $1.28^{+0.53}_{-0.45}$ ~ \cite{ATLAS:WW*_2l2v}& 7(4.5) + 8(20.3)\\
    & & & \\
  \hline 
  & & &\\
   ${\rm GGF ~(CMS)}$  &  $h\to WW^* \to 2\ell2\nu$ & $0.74^{+0.22}_{-0.20}$ ~ \cite{Chatrchyan:2013iaa}& 7(4.9) + 8 (19.4) \\
    & & & \\
  \hline 
  & & &\\
  ${\rm VBF ~(CMS)}$  &  $h\to WW^* \to 2\ell2\nu$ & $0.6^{+0.57}_{-0.46}$ ~ \cite{Chatrchyan:2013iaa}& 7(4.9) + 8 (19.4) \\
  & & &\\
  \hline 
  & & &\\
  $W h\to h\ell\nu ~({\rm CMS})$  &  $h\to WW^* \to 2\ell2\nu$ & $0.56^{+1.27}_{-0.95}$ ~ \cite{Chatrchyan:2013iaa}& 7(4.9) + 8 (19.4) \\ 
  & & &\\
  \hline
  \end{tabular}
    \caption{LHC data used in the global analysis.}\label{tab:LHC-Higgs-Data}
  \end{center}
  \end{table}

To perform the global fit of our CP-odd parameters, we use the standard definition of the chi-square function,
\begin{equation}
\chi^2 = \sum_{X,Y}\frac{(\mu^{X,Y}_{th}-\mu^{X,Y}_{exp})^2}{\Sigma_{X,Y}^2}\label{eq:chisq}   
\end{equation}
where $\mu^{X,Y}_{th}$ is the theoretical signal strength expected in presence of CP-odd operators, and   
$\mu^{X,Y}_{exp}$ is the experimental signal strength reported by the LHC experiments.  
$\Sigma_{X,Y}$ is the experimental uncertainty in $\mu^{X,Y}_{exp}$. The experimental data reported generally has
unsymmetrical uncertainties $\Sigma_{X,Y}^+$ and $\Sigma_{X,Y}^-$. The $\Sigma_{X,Y}$ that we use symmetrizes these uncertainties
through the following definition,
\begin{equation}
 \Sigma_{X,Y} = \sqrt{\frac{(\Sigma_{X,Y}^+)^2 + (\Sigma_{X,Y}^-)^2}{2}}.
\end{equation}
Since the LHC data that we use includes data from both 7 and 8 TeV LHC runs, the 
theoretical signal strength in Eq.(~\ref{eq:chisq}) is obtained after combining 
the signal strengths calculated for 7 and 8 TeV LHC. For that we have used following 
formula~\cite{Corbett:2012dm},
\begin{equation}
\mu^{XY}_{th} = \frac{\mu^{XY}_{th,8} \sigma_8^{\rm SM}\mathcal{L}_8 + 
\mu^{XY}_{th,7} \sigma_7^{\rm SM}\mathcal{L}_7}{\sigma_8^{\rm SM}\mathcal{L}_8 +\sigma_7^{\rm SM}\mathcal{L}_7} 
\end{equation}
where ${\mathcal{L}_7}$ and ${\mathcal{L}_8}$ are the luminosities at 7 and 8 TeV, respectively, and 
$\sigma_7^{\rm SM}$ and $\sigma_8^{\rm SM}$ are the SM cross sections at those energies. These cross 
sections are listed in Table~\ref{tab:smXS}.

\begin{figure}

    \begin{subfigure}[b]{0.3\linewidth}
%     \begin{center}
 \includegraphics[width = 150pt]{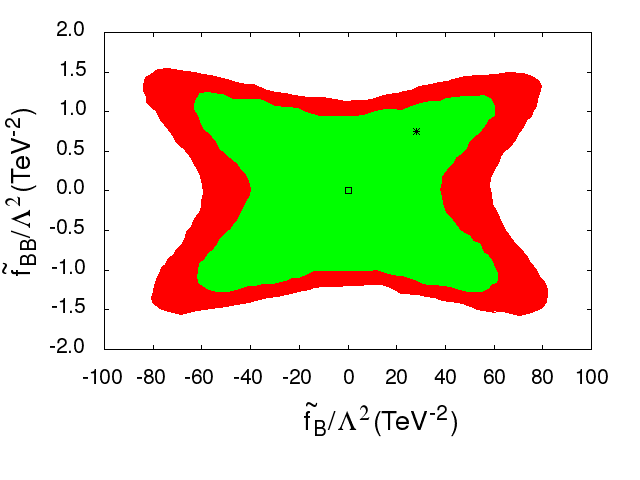}
    \caption{}\label{fig:LHC-2P1}
    
% \end{center}  
   \end{subfigure}
    \begin{subfigure}[b]{0.3\linewidth}
  \includegraphics[width = 150pt]{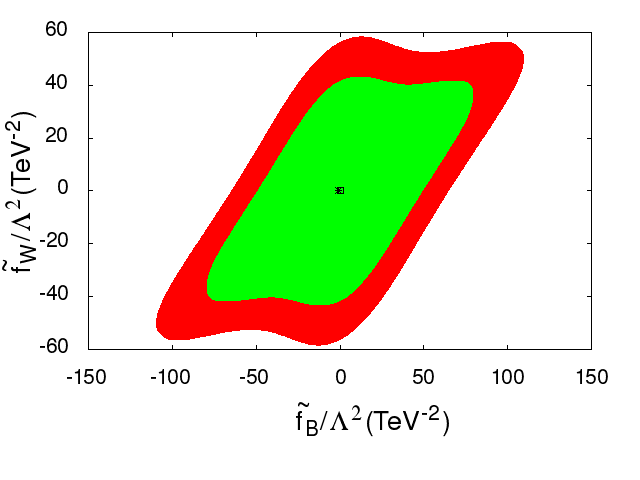}
     \caption{}\label{fig:LHC-2P2}
   
   \end{subfigure}
     \begin{subfigure}[b]{0.3\linewidth}
    \includegraphics[width = 150pt]{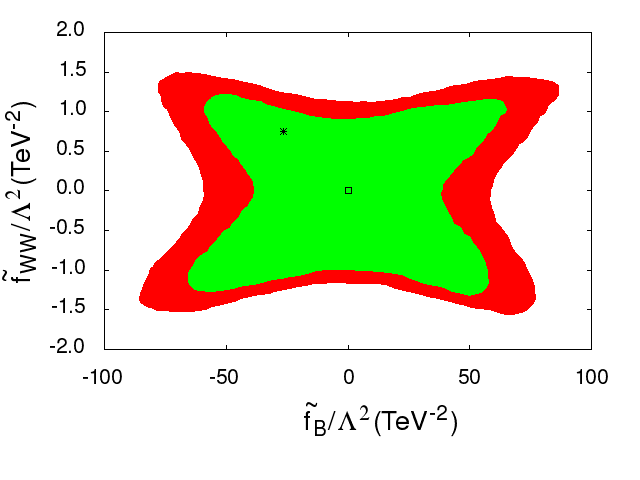}
       \caption{}\label{fig:LHC-2P3}
   
   \end{subfigure}
       \begin{subfigure}[b]{0.3\linewidth}
     \includegraphics[width = 150pt]{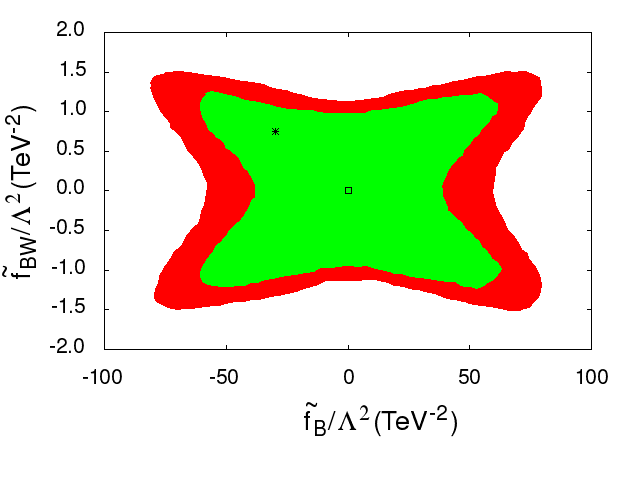}
        \caption{}\label{fig:LHC-2P4}
   
   \end{subfigure}
        \begin{subfigure}[b]{0.3\linewidth}
   \includegraphics[width = 150pt]{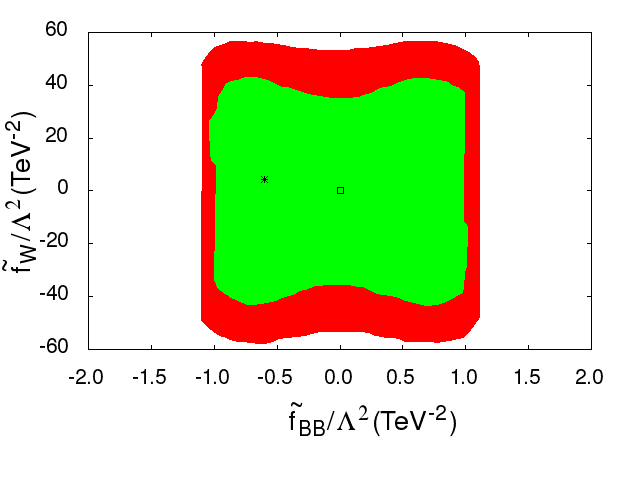}
      \caption{}\label{fig:LHC-2P5}
   
   \end{subfigure}
      \begin{subfigure}[b]{0.3\linewidth}
   \includegraphics[width = 150pt]{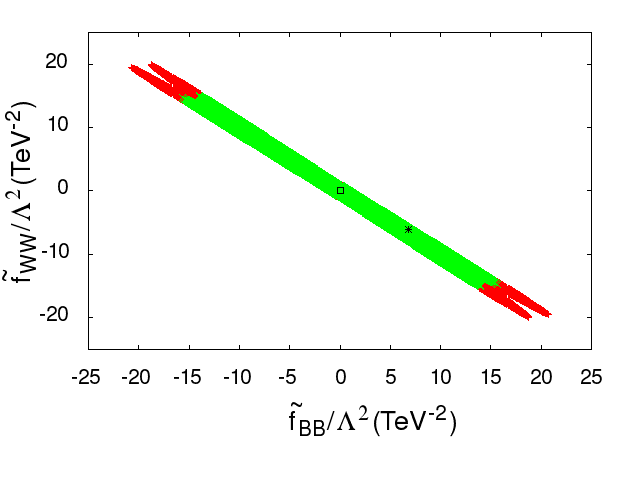}
      \caption{}\label{fig:LHC-2P6}
   
   \end{subfigure}
      \begin{subfigure}[b]{0.3\linewidth}
   \includegraphics[width = 150pt]{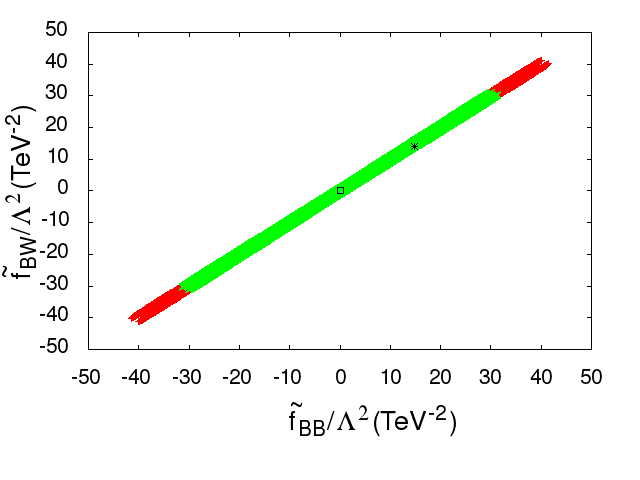}
      \caption{}\label{fig:LHC-2P7}
   
   \end{subfigure}
      \begin{subfigure}[b]{0.3\linewidth}
   \includegraphics[width = 150pt]{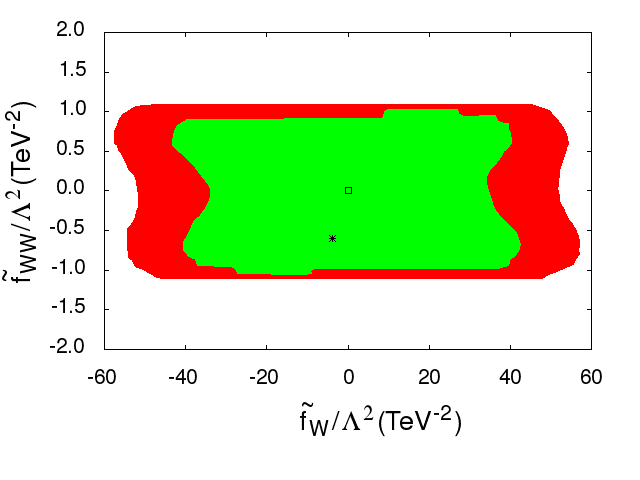}
      \caption{}\label{fig:LHC-2P8}
   
   \end{subfigure}
      \begin{subfigure}[b]{0.3\linewidth}
   \includegraphics[width = 150pt]{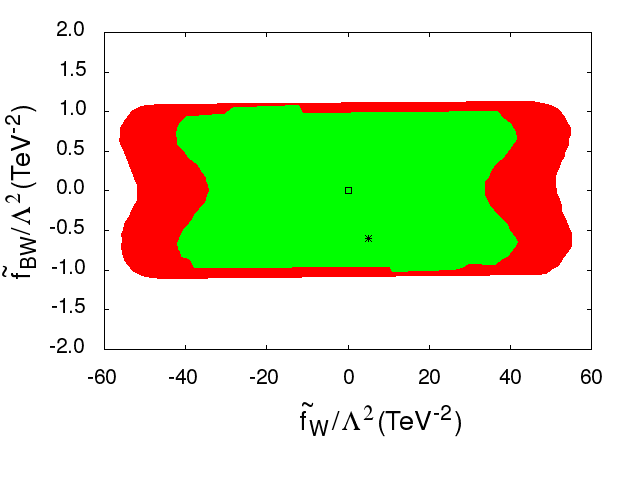}
      \caption{}\label{fig:LHC-2P9}
   
   \end{subfigure}
      \begin{subfigure}[b]{0.3\linewidth}
   \includegraphics[width = 150pt]{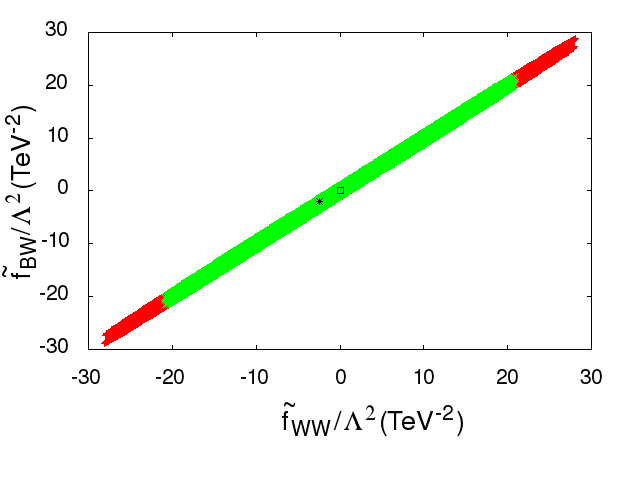}
      \caption{}\label{fig:LHC-2P10}
   
   \end{subfigure}
  \caption{Global fits of the CP-odd parameters keeping two parameters nonzero at a time.
  The point (0,0) corresponds to the SM point and the (*) represents the best fit point. The green region corresponds
  to the 68 percent confidence interval and the red region to the 95 percent confidence interval, respectively. The best
  fit point is doubly degenerate up to a sign flip of the best fit point coordinates.}
  \label{fig:LHC-2P}
 \end{figure}

% \pagebreak
 \begin{figure}
 \begin{subfigure}[b]{0.3\linewidth}
 \includegraphics[width = 150pt]{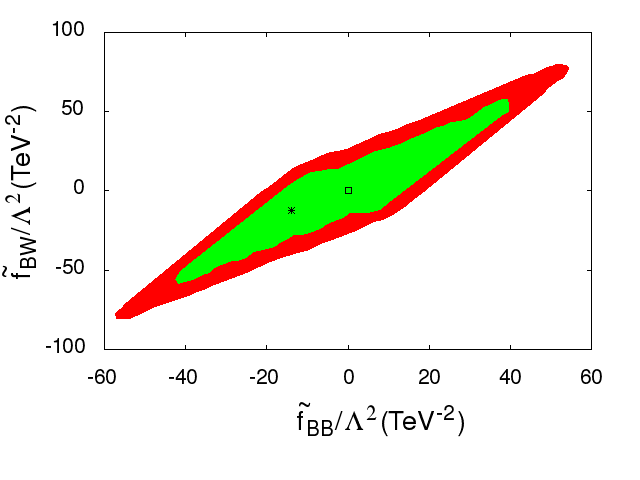}
     \caption{}\label{fig:LHC-3P1}
   
   \end{subfigure}
 \begin{subfigure}[b]{0.3\linewidth}
 \includegraphics[width = 150pt]{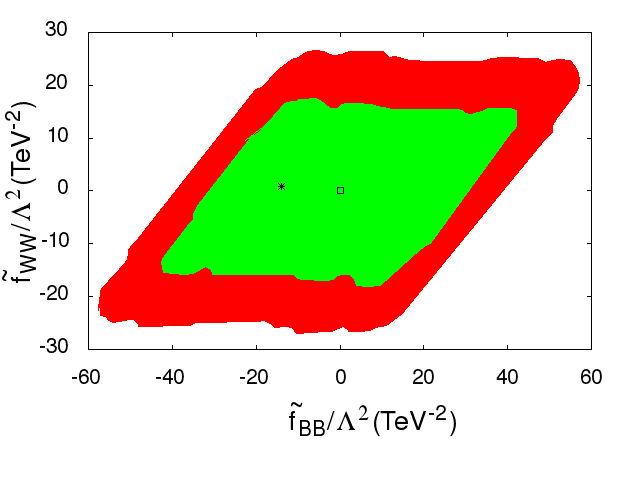}
     \caption{}\label{fig:LHC-3P2}
   
   \end{subfigure}
 \begin{subfigure}[b]{0.3\linewidth}
 \includegraphics[width = 150pt]{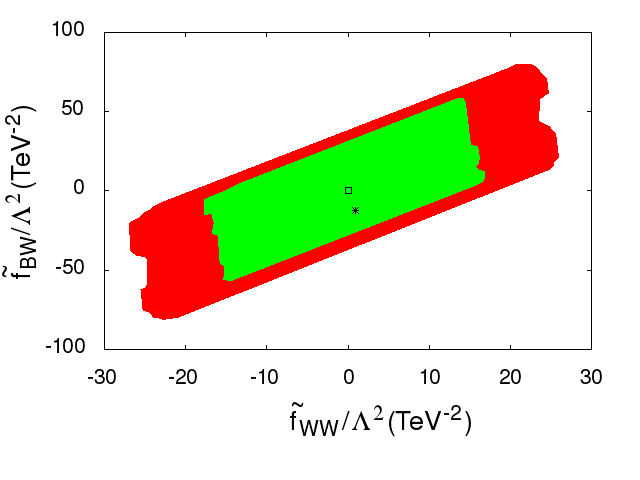}
     \caption{}\label{fig:LHC-3P3}
   
   \end{subfigure}
 \caption{Marginalized global fits of the CP-odd parameters with $\tilde f_W = \tilde f_B = 0$.}
\label{fig:LHC-3P-A}
\end{figure}
 
  \begin{figure}
 \begin{center}
 \begin{subfigure}[b]{0.3\linewidth}
  \includegraphics[width = 150pt]{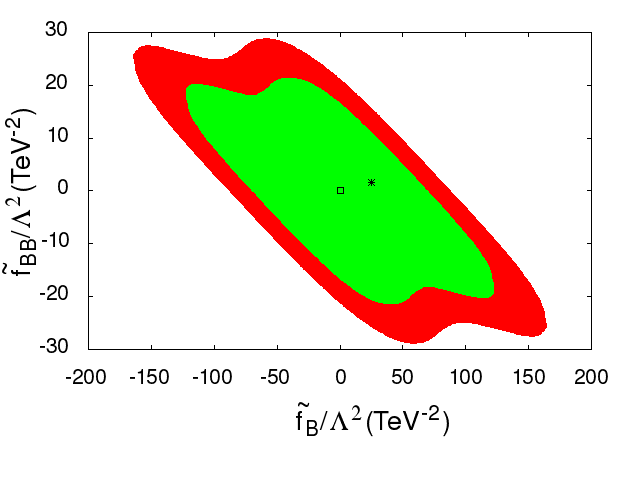}
      \caption{}\label{fig:LHC-3P4}
         \end{subfigure}
  \begin{subfigure}[b]{0.3\linewidth}
 \includegraphics[width = 150pt]{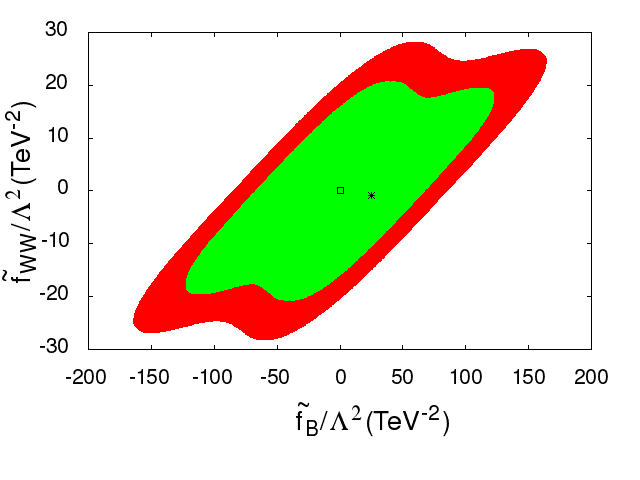}
     \caption{}\label{fig:LHC-3P5}
        \end{subfigure}
 \begin{subfigure}[b]{0.3\linewidth}
 \includegraphics[width = 150pt]{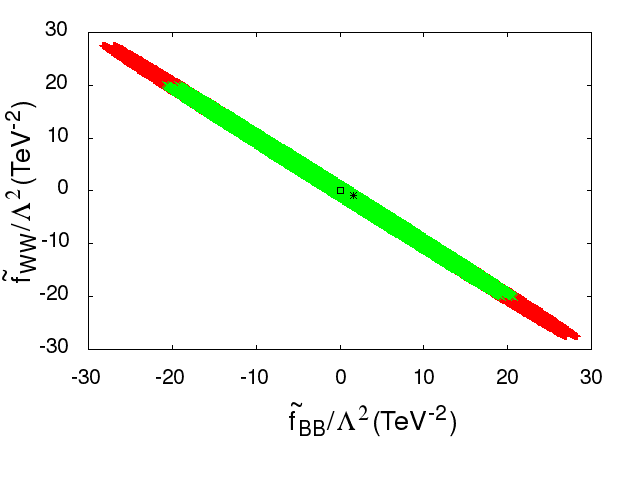}
     \caption{}\label{fig:LHC-3P6}
        \end{subfigure}
 \caption{Marginalized global fits of the CP-odd parameters with  $\tilde f_{W} = \tilde f_{BW} = 0$.}
 \label{fig:LHC-3P-B}
\end{center}

\end{figure}
 
  \begin{figure}
 \begin{center}
   \begin{subfigure}[b]{0.3\linewidth}
  \includegraphics[width = 150pt]{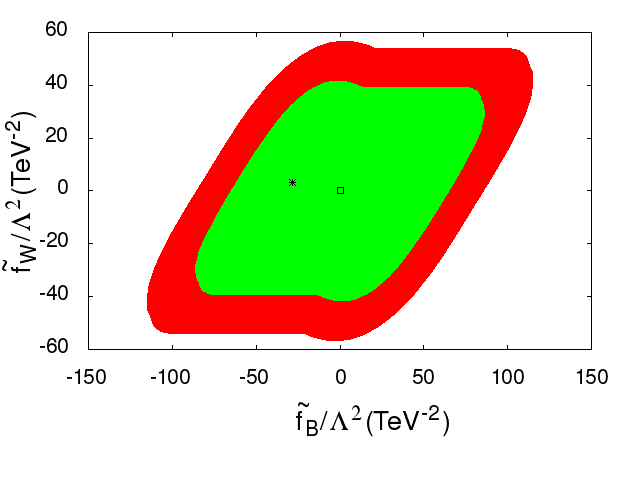}
       \caption{}\label{fig:LHC-3P7}
        \end{subfigure}
          \begin{subfigure}[b]{0.3\linewidth}
 \includegraphics[width = 150pt]{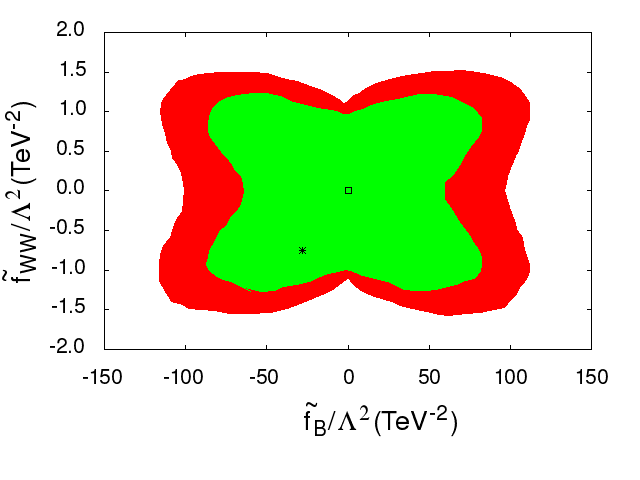}
      \caption{}\label{fig:LHC-3P8}
        \end{subfigure}
          \begin{subfigure}[b]{0.3\linewidth}
 \includegraphics[width = 150pt]{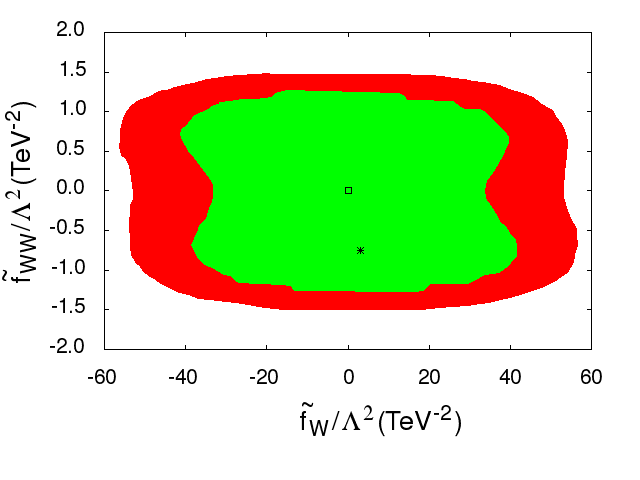}
      \caption{}\label{fig:LHC-3P9}
        \end{subfigure}
\caption{Marginalized global fits of the CP-odd parameters with $\tilde f_{BB} = \tilde f_{BW} = 0$.}
\label{fig:LHC-3P-C}
\end{center}

\end{figure}

In our analysis we have taken total 15 data points which are the most updated ones. We 
have listed them in Table~\ref{tab:LHC-Higgs-Data}. Note that due to large uncertainty 
we do not include Higgs data in $h \to \gamma Z$ decay channel~\cite{Chatrchyan:2013vaa,Aad:2014fia} from CMS and ATLAS.
The global analysis with five CP-odd parameters results into {$ \chi^2_{min}=6.78$}. 
However, in this case the best fit point is very unstable with respect to the step size that 
we choose to scan the parameter space. Also, we find that the $ \chi^2_{min}$ is insensitive to the 
parameters $\tilde f_B$ and $\tilde f_W$. These are the parameters which do not enter the 
$\gamma\gamma h$ vertex and their coefficients are very small compared to those of 
$\tilde f_{BB}, \tilde f_{WW}$ and $\tilde f_{BW}$, which do enter the 
$\gamma\gamma h$ vertex [Eq.(~\ref{eq:aa})]. 
Since the LHC observables have an overall cutoff scale dependence, the ratio $\tilde f_i/\Lambda^2$ can be 
taken as the effective parameter to be constrained. In other words, the constraints from global analysis 
can be easily predicted for any value of $\Lambda$ of interest. 
% {\cgreen Moved here from the end of this section.}}

We organize the constraints on CP-odd 
parameters from global fit of LHC data in the following two parts. 
\begin{itemize}
 \item 
  We have five CP-odd parameters $({\tilde f}_W, {\tilde f}_B, 
{\tilde f}_{BB}, {\tilde f}_{BW}, {\tilde f}_{WW})$ and in the first case we 
consider any two of them to be non-zero and put limits on them.
 There are a total of ten such combinations. 
The allowed parameter space with 68$\%$ and 95$\%$ confidence levels are shown in 
Fig.\ref{fig:LHC-2P}. In generating these plots we have varied parameters freely. 
Even if one considers the perturbativity argument, the tightest 
upper bound on these CP-odd parameters is of the order $\sim 100$.
We can see that in all cases the allowed parameter space is bounded. The 
parameters which enter the $\gamma\gamma h$ vertex ($\gamma\gamma h$ family) {\it i.e.}, $\tilde f_{BB}, \tilde f_{WW}$
and $\tilde f_{BW}$ are in general more constrained than those which do not enter the $\gamma\gamma h$ vertex 
(non-$\gamma\gamma h$ family) {\it i.e.}, $\tilde f_B$ and $\tilde f_W$.  
When any of the three $\gamma\gamma h$ parameters is taken together with $\tilde f_B$ or 
$\tilde f_W$ as shown in Figs~\ref{fig:LHC-2P1},\ref{fig:LHC-2P3},\ref{fig:LHC-2P4},\ref{fig:LHC-2P5},
\ref{fig:LHC-2P8} and \ref{fig:LHC-2P9}, we find that it is much tightly constrained and allowed values are of ${\cal O}(1)$. 
However, among themselves these parameters are highly correlated and cancellation among them leads 
to larger allowed values of ${\cal O}(10)$ (see Figs.~\ref{fig:LHC-2P6},\ref{fig:LHC-2P7} 
and \ref{fig:LHC-2P10}). The nature 
of slope in these figures is related to the relative sign among $\tilde f_{BB}, \tilde f_{WW}$
and $\tilde f_{BW}$. We also note that out of $\tilde f_B$ and $\tilde f_W$, $\tilde f_W$ is 
always more constrained. For example, the maximum allowed value for $\tilde f_B$ is $\sim 100$ 
while the allowed values of $\tilde f_W$ are less than 60, see Fig.~\ref{fig:LHC-2P2}. This observation can be attributed 
to the relative size of their coefficients in various observables. We have also found that the inclination of the 
plot in Fig.~\ref{fig:LHC-2P2} is governed by the $C_{\gamma Zh}$ which enters $Zh$ and VBF production channels and affects 
the $S_{total}$ [Eq.(~\ref{eq:S-total})].

\item In the second case we consider three parameters at a time in the global analyses. Once again there are ten such 
combinations. Following the general conclusions of two-parameter case, we can categorize these combinations into three groups. 
This categorization is based on the number of parameters from the $\gamma\gamma h$ family being present in each combination. Thus we have 
one combination where all the parameters are from $\gamma\gamma h$ family (G1), six combinations where two parameters 
are from $\gamma\gamma h$ family and one from non-$\gamma\gamma h$ family (G2) and three combinations where one is 
from $\gamma\gamma h$ family and the other two are $\tilde f_B$ and $\tilde f_W$ (G3). We present results for three 
representative combinations (one from each group): 
(i) \{ $\tilde f_{BB}, \tilde f_{WW}, \tilde f_{BW}$ \}; 
(ii) \{ $\tilde f_{BB}, \tilde f_{WW}, \tilde f_B$ \};
(iii) \{ $\tilde f_{WW}, \tilde f_B, \tilde f_W$ \}. 
The allowed parameter space for these combinations are shown in Figs.~\ref{fig:LHC-3P-A}, \ref{fig:LHC-3P-B} 
and \ref{fig:LHC-3P-C}, respectively. The two-parameter plots here are obtained after marginalizing over the 
third parameter.

As compared to two-parameter plots (see Fig. \ref{fig:LHC-2P}), 
the allowed region here is more diffused due to the presence of the third parameter. This is particularly 
noticeable for the first set of plots in Fig.~\ref{fig:LHC-3P-A}. We find that $\tilde f_{WW}$ being present in all 
$VVh$ couplings gets stronger bounds, while $\tilde f_{BW}$ is least constrained in set (i). 
From the plots of Fig.~\ref{fig:LHC-3P-B} which belong to set (ii) we can infer that the parameters 
of $\gamma\gamma h$ family are more constrained. However, mutual cancellation still allows values of ${\cal O}(30)$ 
for them. 
The opposite inclinations of the plots in ~\ref{fig:LHC-3P4} and ~\ref{fig:LHC-3P5} can be related to how 
$\tilde f_{BB}$ and  $\tilde f_{WW}$ enter in $\gamma\gamma h$ vertex to maintain cancellation and it is confirmed 
by the nature of the slope in Fig.~\ref{fig:LHC-3P6}.
Note that Fig.~\ref{fig:LHC-3P6} is very much similar to the corresponding plot in Fig.~\ref{fig:LHC-2P6}. 
This is expected because the global analysis is not very sensitive to $\tilde f_B$. 
Interestingly, the parameter sets \{$\tilde f_{WW},\tilde f_{BW},\tilde f_{W}$\} and 
\{$\tilde f_{BB},\tilde f_{BW},\tilde f_{B}$\} which would also belong to group G2
are less constrained. We find that the parameter regions are still bounded but boundary values for $\tilde f_W$ and 
$\tilde f_B$ in these sets are about 800 and 2500, respectively.
For the parameters in set (iii) we can conclude that  $\tilde f_{WW}$ being the only parameter 
present from the $\gamma\gamma h$ family is very tightly constrained as shown by the plots in Figs.~\ref{fig:LHC-3P8}
and \ref{fig:LHC-3P9}. 

\end{itemize}

We have also checked that the constraints on CP-odd parameters from the direct and indirect measurements of the Higgs 
total width~\cite{Chatrchyan:2013mxa,Khachatryan:2014iha} are weaker than those obtained from the global analysis.
 On the whole, $\tilde f_{BB}$, $\tilde f_{WW}$ and $\tilde f_{BW}$ contribute to the $\gamma \gamma h$ vertex at 
tree level and thus are constrained rather tightly compared to $\tilde f_{B}$ and $\tilde f_{W}$. The fact that one has SM contribution at the one-loop level only is responsible
for this. $\tilde f_{B}$ and $\tilde f_{W}$ are relatively loosely constrained due to the lack of sufficient data on the 
channel $h\rightarrow \gamma Z$, to which they contribute.

 The constraints on the parameter space of our CP-violating operators, as obtained from
 global fits of the (7 + 8) TeV data, are expected to be improved in the high energy runs. A tentative 
 estimate of such improvements, as also that in a linear $e^+ e^-$ collider, can be found, for example, in
\cite{Englert:2014uua}. Going by the estimates of \cite{Englert:2014uua}, the uncertainty in the signal 
strength measurements in the next run can be reduced to 33$\%$ of the present uncertainties as
obtained by both ATLAS and CMS for the $\gamma\gamma$ and $WW^*$ final states in the gluon fusion channel.
A precise answer on the improvement of our limits, however, depends also on any possible shift in the 
central values of the measured signal strengths in different channels. This in turn is also a function
of the various systematic uncertainties in the new run, and therefore we have to wait for more data
before some precise conclusions can be drawn.
A similar consideration applies to a linear collider; precision in coupling measurements down to 1$\%$ 
is expected there in principle \cite{Englert:2014uua}, but the available statistics as well as the systematics 
need to be known before concrete estimates emerge.

\section{Constraints from EDMs}

\begin{figure}[t]
\begin{center}
\includegraphics[width = 0.8\linewidth]{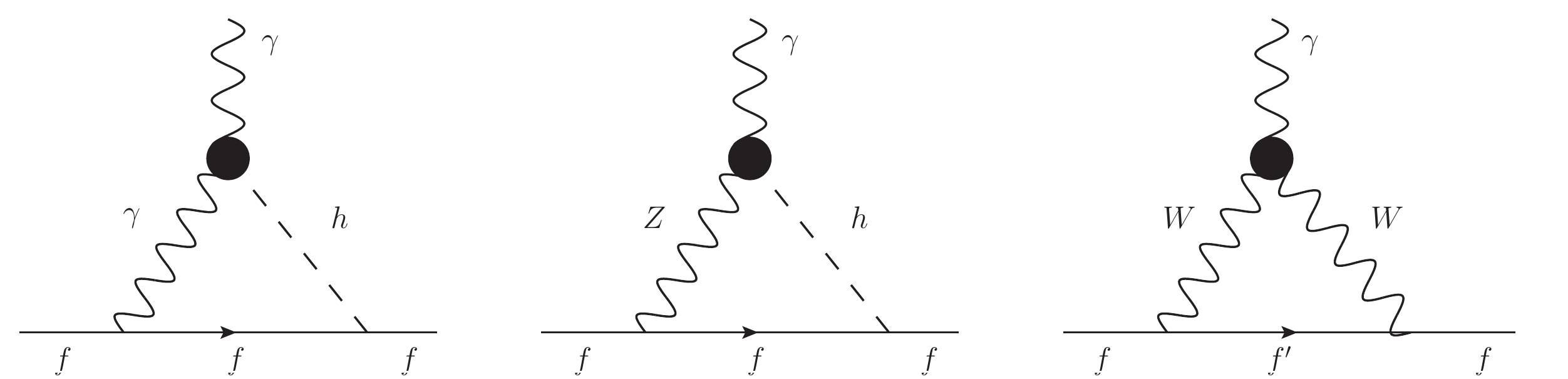}
\caption{One-loop diagrams contributing to fermion EDMs. The blobs show the effective vertices arising out of the
CP-odd operators.}\label{fig:EDM-1Loop}
\end{center}
\end{figure}

The fermionic electric dipole moment receive an additional contribution
from these new CP-odd higher-dimensional operators involving Higgs and pair
of gauge bosons. Nonobservation of any fermionic EDMs puts severe 
constraints on the parameters ${\tilde f_i}$s. 
The fermion EDM operator is defined as, 
\begin{eqnarray}
 -\frac{1}{2}\; d_f \; {\bar \psi}(p_2)\; i \gamma^5 \sigma^{\mu \nu}\; \psi(p_1)\: F_{\mu\nu},
\end{eqnarray}
where, $\sigma^{\mu\nu} = \frac{i}{2} [ \gamma^\mu,\gamma^\nu ]$ and $d_f$ is known as the fermion EDM form factor. 
Nonvanishing EDMs provide clear hint of 
CP-violation~\cite{Pospelov:2005pr,Engel:2013lsa}. In the standard model, CP violation occurs due to quark mixing and it is quite weak 
(1 part in 1000)~\cite{Agashe:2014kda}. On top of that 
the first nonzero contribution to EDM operator in the SM appears at three loop level in quark sector, while, for 
leptons it arises at four loop level. The present upper limits on electron and neutron EDMs are much larger than
the values predicted by the SM~\cite{Czarnecki:1997bu}.
In presence of CP-odd gauge-Higgs operators $\gamma\gamma h$, $\gamma Zh$  and $WW\gamma$ 
couplings are modified, because of which the leading contribution to 
fermion EDMs appears at one-loop level. Due to this the fermion EDM measurements can provide stringent bounds on the 
CP-odd parameters. Note that the contribution to the fermion EDMs from CP-odd $WWh$ and $ZZh$ vertices 
can result only at two-loop level. Since the two-loop effects are expected to be subdominant, we will
derive constraints on the CP-odd parameters from one-loop fermion EDM calculations.
The diagrams that contribute to the fermion EDM at one-loop level are shown in Fig.\ref{fig:EDM-1Loop}.

The expression for the fermion EDM form factor $d_f$ at one-loop due to the $\gamma\gamma h$, $\gamma Zh$  and $WW\gamma$
vertices is given by the following equation\footnote{We have observed a relative sign change in the contribution 
of the $WW\gamma$ diagram to the EDM, for the $u$ and $d$ quarks, which is taken care of by the factor $I_f$ in $\tilde a_3$. 
It was unaccounted by Ref.~\cite{Chang:2013cia} where a similar calculation is done.}:
\begin{eqnarray}\label{1loop-edm}
d_f~=~\frac{m_fe\alpha}{\pi v^2}\Big[\tilde{a}_1K_1(\Lambda,m_h)+
\tilde{a}_2K_2(\Lambda,m_Z,m_h)+
\tilde{a}_{3}K_1(\Lambda,m_W)\Big]
\end{eqnarray}
where,  
\begin{eqnarray}
\tilde{a}_1= -\frac{Q_f}{4 s_W^2} C_{\gamma\gamma h};~~
\tilde{a}_2= \frac{\Big(\frac{1}{2}I_f-Q_f s_{W}^2\Big)}{t_W s_{2W}^2} C_{\gamma Zh};~~
\tilde{a}_{3}=-\frac{I_{f}}{4 s_W^3} C_{WW\gamma}.
\end{eqnarray}
Here, $v$ is the electroweak symmetry breaking scale, $\alpha$ is the fine structure constant, $I_f$ is the third component 
of the fermion Isospin and $Q_f$ is its electric charge quantum number. We have 
neglected the fermion masses ($m_f$) with respect to other mass scales in the loop.
The one-loop factors $K_1$ and $K_2$ are calculated in dimensional regularization ($d=4-2\epsilon$). Since these loops are 
UV divergent, we renormalize them in $\overline{\rm MS}$ scheme and identify the renormalization scale with the  
cutoff $\Lambda$. The expressions for $K_1$ and $K_2$ are given by,
\begin{eqnarray}
K_1(\Lambda,x)~&=&~\frac{v^2}{\Lambda^2}\Big[\frac{1}{2}{\rm ln}\frac{\Lambda^2}{x^2} + \frac{3}{4} \Big],\\
K_2(\Lambda,x,y)~&=&~\frac{v^2}{\Lambda^2}\Big[ \frac{1}{2}\frac{x^2{\rm ln}\frac{\Lambda^2}{x^2}-y^2{\rm ln}\frac{\Lambda^2}{y^2}}{x^2-y^2} + \frac{3}{4} \Big].
\end{eqnarray}
In the above, the finite factors of $\frac{3}{4}$ are artifact of dimensional regularization. 
 These factors do not appear in naive cutoff regularization. 

 The latest experimental bounds on the electron and neutron EDMs are~\cite{Baron:2013eja,Beringer:1900zz,Baker:2006ts}, 
\begin{eqnarray}
 |d_e| &<& 8.7 \times 10^{-29}\; e \;{\rm cm} \nn \\
 |d_n| &<& 2.9 \times 10^{-26}\; e \;{\rm cm}.
\end{eqnarray}
Note that the EDM contribution is ${\cal O}(\frac{1}{\Lambda^2})$ in the cutoff scale, therefore, it 
is expected to provide stronger bounds on CP-odd parameters than those obtained 
from EWP and LHC data. 
Like the electroweak precision observables calculated in section 3, 
the EDMs also have explicit dependence on $\Lambda$. 
Due to this, we provide
EDM constraint equations for three different choices of cutoff scale $\Lambda$ = 1, 5 and 10 TeV. 
\begin{itemize}
 \item $\Lambda$ = 1 TeV 
\begin{eqnarray}
 |d_e| &\equiv& |233.86 \tilde f_{B} + 260.45 \tilde f_{W} - 390.92 \tilde f_{BB} - 337.72 \tilde f_{WW} + 858.63 \tilde f_{BW}  |< 1 \label{e_edm 1 TeV} \nn \\
 |d_n| &\equiv& |  7.02 \tilde f_{B} +  13.81 \tilde f_{W} -   8.91 \tilde f_{BB} +   4.66 \tilde f_{WW} +  22.96 \tilde f_{BW}  |< 1\label{n_edm 1 TeV}
\end{eqnarray}

 \item $\Lambda$ = 5 TeV 
\begin{eqnarray}
 |d_e| &\equiv& |13.87 \tilde f_{B} + 15.63 \tilde f_{W} - 24.60\tilde f_{BB} - 21.08 \tilde f_{WW} + 52.34 \tilde f_{BW}  |< 1 \label{e_edm 5 TeV}\nn \\
 |d_n| &\equiv& | 0.40 \tilde f_{B} +  0.85 \tilde f_{W} -  0.57\tilde f_{BB} +  0.33 \tilde f_{WW} +  1.36 \tilde f_{BW}  |< 1\label{n_edm 5 TeV}
\end{eqnarray}

  \item $\Lambda$ = 10 TeV 
 \begin{eqnarray}
 |d_e| &\equiv& |3.95\tilde f_{B} + 4.47\tilde f_{W} - 7.11\tilde f_{BB} - 6.08\tilde f_{WW} + 15.02\tilde f_{BW}  | <1 \label{e_edm 10 TeV} \nn \\
 |d_n| &\equiv& |0.11\tilde f_{B} + 0.24\tilde f_{W} - 0.16\tilde f_{BB} + 0.10\tilde f_{WW} +  0.39\tilde f_{BW}  | <1\label{n_edm 10 TeV}
 \end{eqnarray}
\end{itemize}
The neutron EDM form factor ($d_n$) is calculated in terms of constituent quark EDMs using the relation 
$d_n~=~ \frac{4}{3}d_d-\frac{1}{3}d_u$, from the chiral-quark model~\cite{Dib:2006hk}.
In calculating the above constraints on the EDMs we  take
 $\alpha~=~1/137$  and $M_H = 126\; {\rm GeV}$. Because of a stronger experimental limit on electron 
 EDM, the coefficients of parameters in $d_e$ are larger than those in $d_n$. Our constraint equations for electron EDM 
 form factor differ by an order of magnitude from those obtained in Ref.~\cite{Chang:2013cia} mainly because we have used the most updated experimental 
 bound on electron EDM~\cite{Baron:2013eja}.

 \begin{figure}
%  \begin{center}
 \includegraphics[width = 0.30\linewidth]{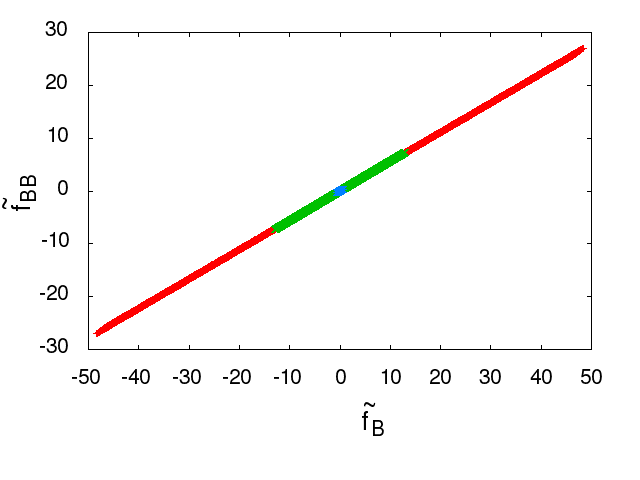}
 \includegraphics[width = 0.30\linewidth]{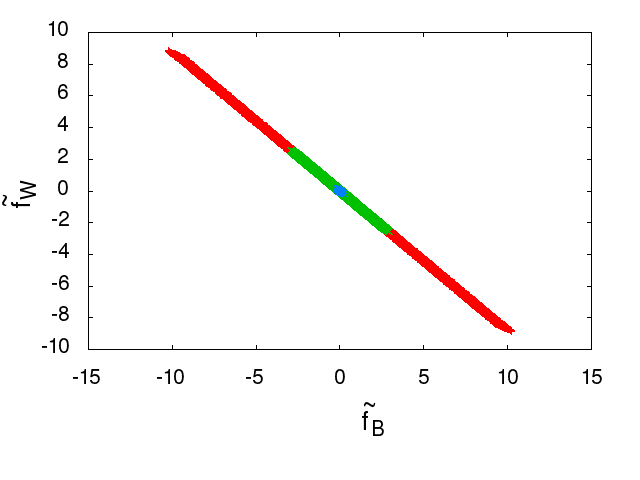}
 \includegraphics[width = 0.30\linewidth]{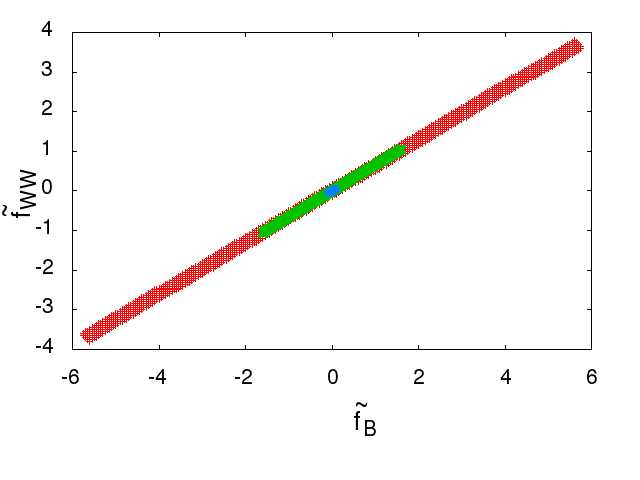}
 \includegraphics[width = 0.30\linewidth]{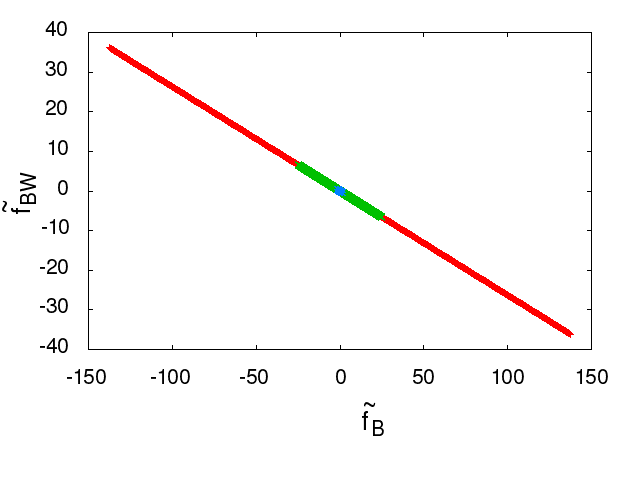}
  \includegraphics[width = 0.30\linewidth]{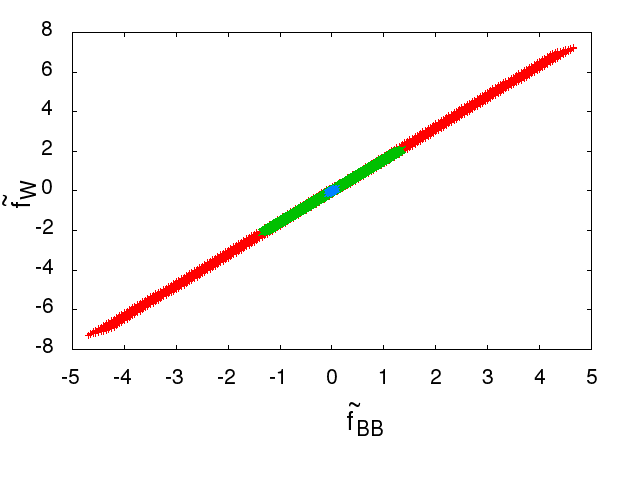}
 \includegraphics[width = 0.30\linewidth]{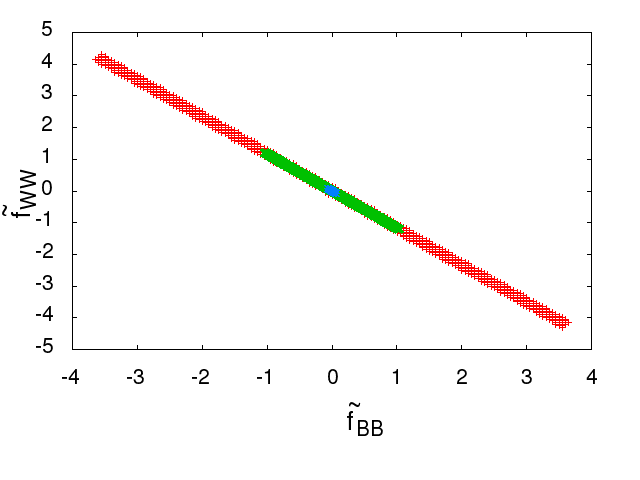}
 \includegraphics[width = 0.30\linewidth]{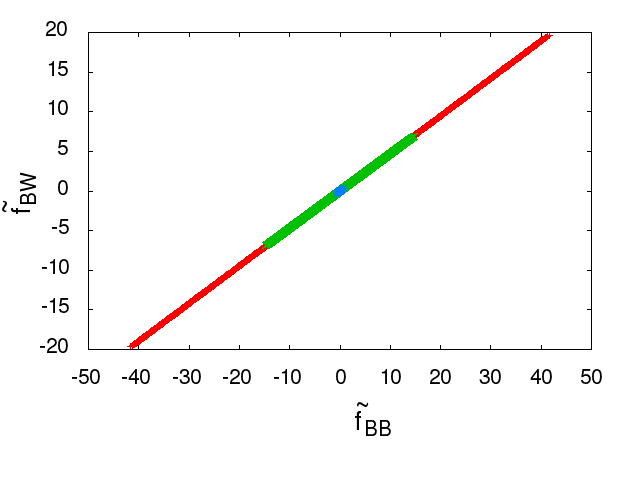}
 \includegraphics[width = 0.30\linewidth]{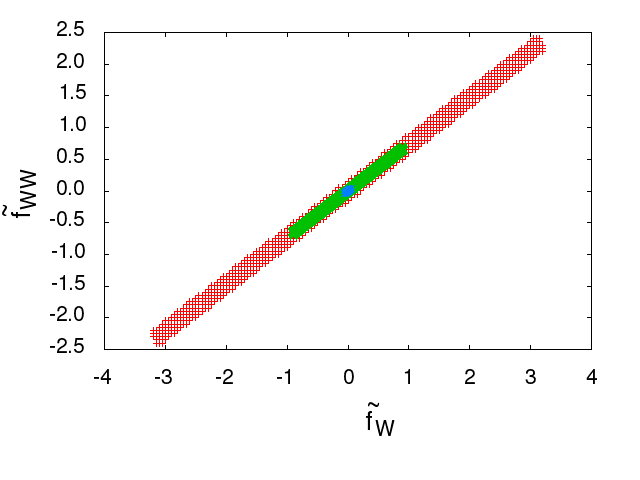}
 \includegraphics[width = 0.30\linewidth]{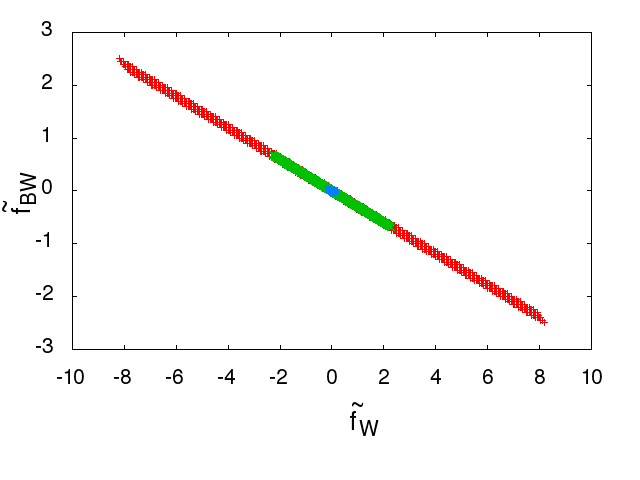}
 \includegraphics[width = 0.30\linewidth]{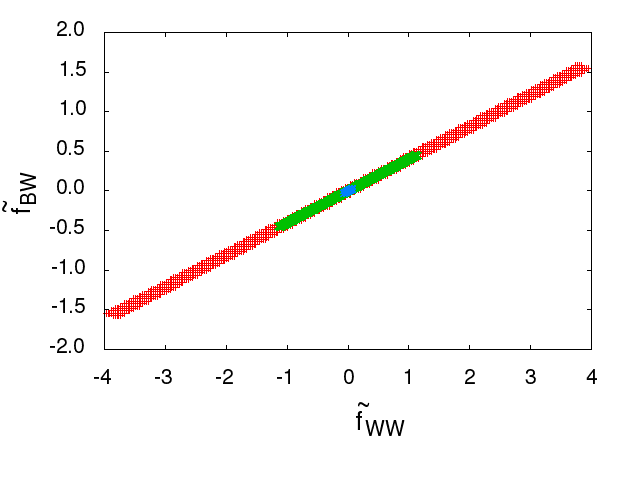}
 \caption{EDM constraints keeping two parameters nonzero at a time 
 for three representative values of $\Lambda = 1({\rm blue}), 5({\rm green})~ \&~ 10({\rm red}) $ TeV.}\label{fig:EDM-2P}
%  \end{center}
  \end{figure}

   \begin{figure}
%  \begin{center}
\begin{subfigure}[b]{1.0\linewidth}
 \includegraphics[width = 0.30\linewidth]{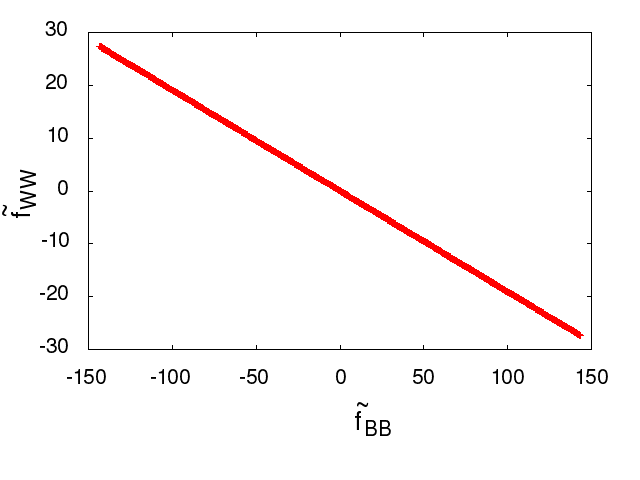}
 \includegraphics[width = 0.30\linewidth]{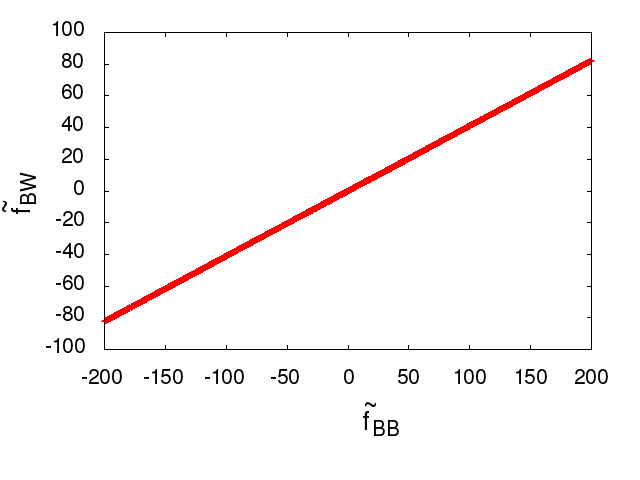}
 \includegraphics[width = 0.30\linewidth]{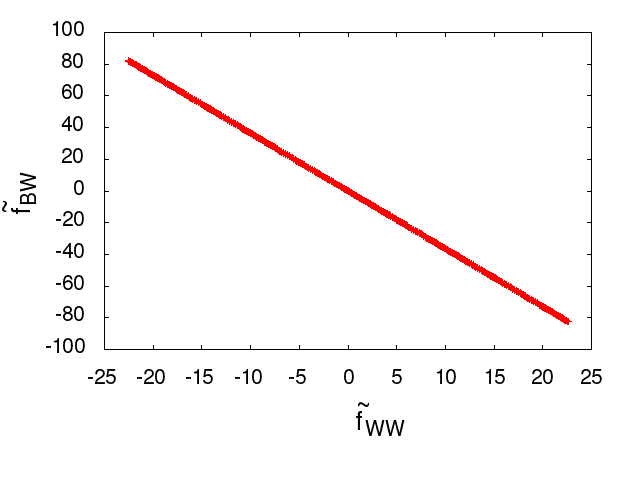}
 \caption{}\label{fig:EDM-3P1}
 \end{subfigure}
 
 \begin{subfigure}[b]{1.0\linewidth}
 \includegraphics[width = 0.30\linewidth]{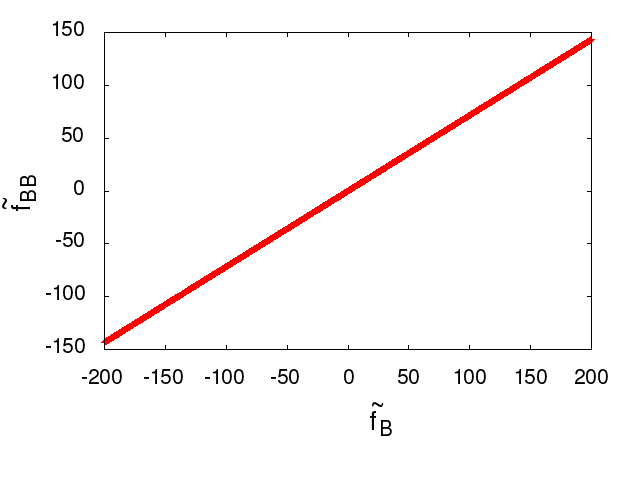}
  \includegraphics[width = 0.30\linewidth]{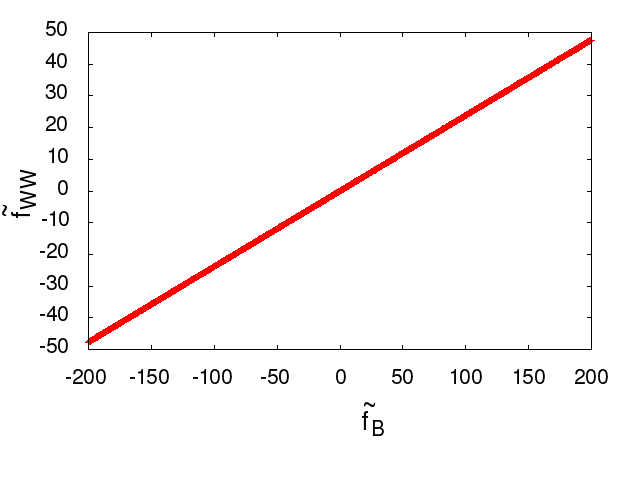}
 \includegraphics[width = 0.30\linewidth]{./fbb-vs-fww-edm-3.png}
 \caption{}\label{fig:EDM-3P2}
  \end{subfigure}
 
 \begin{subfigure}[b]{1.0\linewidth}
 \includegraphics[width = 0.30\linewidth]{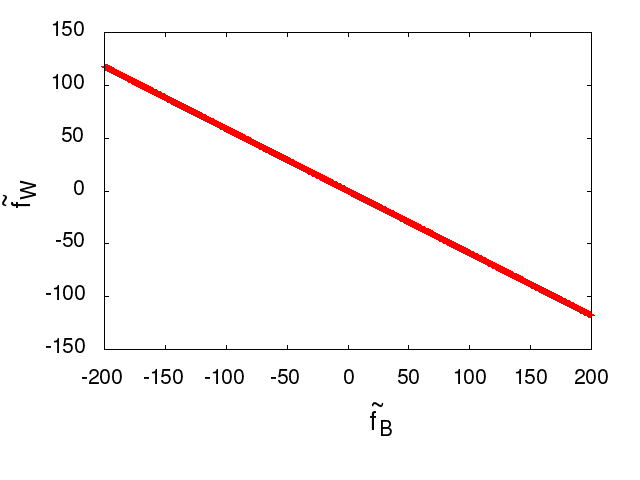}
 \includegraphics[width = 0.30\linewidth]{./fb-vs-fww-edm-3.png}
 \includegraphics[width = 0.30\linewidth]{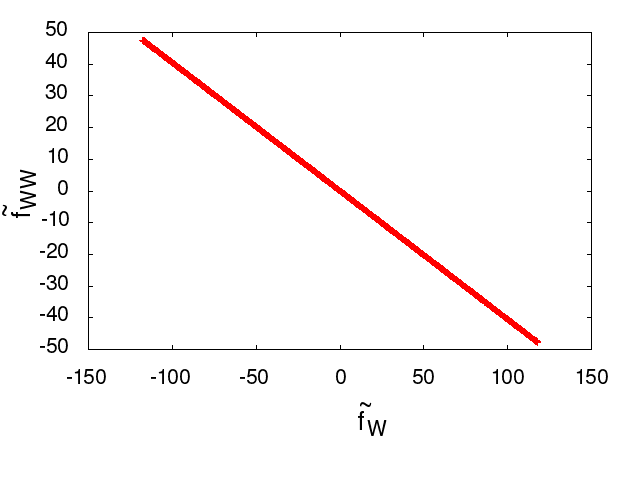}
 \caption{}\label{fig:EDM-3P3}
  \end{subfigure}
 \caption{EDM constraints  
 with (a)$\tilde f_W = \tilde f_B = 0$, (b)$\tilde f_{W} = \tilde f_{BW} = 0$ 
and (c)$\tilde f_{BB} = \tilde f_{BW} = 0$
 for $\Lambda = 1 $ TeV. Parameters are varied in the range between -200 to 200.}\label{fig:EDM-3P}
%  \end{center}
  \end{figure}

We first consider the case when any two of the five parameters are kept nonzero. 
In this case, we freely vary the parameters. The constraints on the parameters are 
shown in Fig~\ref{fig:EDM-2P} for $\Lambda = 1,5 $ and 10 TeV respectively. In all 
combinations we get bounded regions.
The inclinations of the constraint regions 
can be understood from the relative sign between the parameters in the expressions for the EDMs.
As expected the constraints for $\Lambda=1$ TeV 
are tighter than those for $\Lambda =$ 5 and 10 TeV and this is corroborated by the size of the coefficients 
entering in EDM expressions. In $\Lambda=1$ TeV case, the allowed values for parameters 
can reach ${\cal O}(1)$ values at maximum. As we push the cutoff scale higher, the allowed range for parameters 
also increases. For example, for $\Lambda=10$ TeV the allowed values can become  ${\cal O}(10)$ or larger in some
cases.

A naive comparison with the EDM calculation carried out in Ref.~\cite{McKeen:2012av} with only 
$\gamma\gamma h$ CP-odd coupling suggests that for $\Lambda=1~{\rm TeV}$, 
the constraint equation for electron EDM measurement would be $|\tilde f_{BB} + \tilde f_{WW}| \lesssim 0.0036$. 
In the presence of CP-odd $\gamma Zh$ coupling, which arises from same operators, this constraint equation would change. 
For example, our calculation for electron EDM constraint implies, $|\tilde f_{BB} + 0.86 \tilde f_{WW}| \lesssim 0.0026$.
In both the cases, the parameters would be allowed to take very large but fine-tuned values. 
However, we would like to point out that after including the constraints from the neutron EDM measurement, these 
parameters cannot take values larger than ${\cal O}(0.1)$.

When we take three parameters nonzero at a time, the parameters are scanned in the range -200 to 200. 
In Fig~\ref{fig:EDM-3P}, we give two dimensional projection plots of the three dimensional constraint region for $\Lambda=1$ 
TeV. For comparison purpose, we present the plots in three categories discussed in the global analysis.
We can see that with three parameters present, the constraints are more relaxed than when only two of them 
are nonzero. However, the parameters are still quite correlated. 
$\tilde f_B$ and $\tilde f_{BB}$ can often reach the boundary of the scanned regions.
In fact it is very difficult to obtain closed boundaries. As we increase the range for parameter scan 
the allowed values for CP-odd parameters become very large [${\cal O}(1000)$]. However, it is important to 
note that for too large values of parameters, the two-loop EDM constraints may become relevant and, 
therefore, should also be taken into account. 
As we turn on more parameters, the correlation among parameters constrained by EDM  is relaxed and 
the allowed parameter space also expands.\\
{ From the experimental perspective, a number of new EDM experiments promise to improve the level of sensitivity 
by one to two orders of magnitude in the coming years. For example, The Institut Laue Langevin (ILL) cryogenic experiment and the Spallation Neutron Source (SNS) 
nEDM experiment (\cite{Peng:2008ha}, \cite{Ito:2007xd}, \cite{Raidal:2008jk}) aim at improving the upper limit on neutron EDM by two orders 
of magnitude, i.e down to $~{\cal O}(10^{-28}) e.cm$. 
% This would lead to further tightening of the EDM constraints on the allowed parameter space. 
This would imply that the numerical coefficients in the constraint equations for $d_n$ in Eqs. (\ref{n_edm 1 TeV}), (\ref{n_edm 5 TeV}) and (\ref{n_edm 10 TeV}) 
would become stronger by almost two orders of magnitude and thus the allowed parameter space for $\tilde{f}'$s will be even more severely constrained, 
unless, of course, there is direct evidence of neutron EDM in the aforesaid experiments.

% Looking at the EDM constraint equations, one can also argue that as the future experiments are going to
% improve the bounds on the upper limits of the electron and neutron EDMs, the coefficients of these equations
% are going to become larger in magnitude, resulting in even tighter constraints on the allowed parameter values.
}

\section{Discussion}
We now highlight the important features of our analysis presented in sections 3, 4 and 5. We try
to draw a comparative picture and address some of the issues relevant to the analysis.

% % We now highlight the important features of our analysis presented in sections 3, 4 and 5 and 
% try to draw a comparative picture. 

\begin{itemize}

\item
In the two parameter case, among all the constraints, the EWP constraints on CP-odd parameters are always the weakest 
while EDM constraints are the strongest. In general, the correlation among parameters is stronger in EWP and EDM cases 
as compared to the LHC case.

\item 
In case of global fit of LHC data whatever contributes to $h \rightarrow \gamma \gamma$ receives stronger constraints, 
 since here the ``tree level'' contributions from the dimension six operators are essentially at the same level as 
 the ``one-loop'' SM contributions. There are three such parameters namely $\tilde f_{BB}, \tilde f_{WW}$ and $\tilde f_{BW}$. 
{ Only a strong cancellation or correlation can keep the individual values of 
 these parameters large. }
 
  \item The same should apply to parameters contributing to $h \rightarrow \gamma Z$ channel which also includes $\tilde f_B$ and $\tilde f_W$. 
However the limits based on current data are rather weak there and we have not included them in our analysis. The global analysis 
puts some limits on $\tilde f_B$ and $\tilde f_W$ as the $\gamma Zh$ coupling modifies the $Zh$ and VBF production channels and the total 
Higgs decay width.
 
 \item
{ The channels dependent on the $WWh$ and $ZZh$ couplings 
yield relatively weak constraints from the global fits as a whole, since the higher dimensional interactions are 
inadequate to override the tree level SM contributions.}

\item
In two parameter case, LHC data bounds on $\tilde f_{BB}, \tilde f_{BW}$ and $\tilde f_{WW}$ 
 are stronger when any of these is combined with $\tilde f_B$ or $\tilde f_W$. These bounds are 
 comparable to corresponding bounds obtained from EDM measurements. 
 
 \item In three parameter case, the bounds from LHC data are stronger than 
 those obtained from EDMs. In fact, for some parameters (for example, $\tilde f_B$) values 
 of the order 1000 are also allowed by EDM data.
 However, the parameters constrained from EDMs still 
 display a tight correlation. 
 
 \item

If we allow dimension-6 CP-even operators to coexist with the CP-odd ones, the EWP and LHC 
observables would receive contributions at  ${\cal O}(1/\Lambda^2)$ (as a result of interference with the SM) 
as well as at ${\cal O}(1/\Lambda^4)$. 
Since there is no interference between CP-odd and CP-even operators in total rates, the CP-odd interactions studied
by us should contribute to signal strengths on the order  of  $(1/\Lambda^4)$ .
For the sake of consistency, at  ${\cal O}(1/\Lambda^4)$ 
the contribution from dimension-8 CP-even operators via its interference with the SM should also be considered. 
Thus, in the presence of CP-even operators, the constraints from LHC and precision data can be relaxed. However,  
under the assumption that the effect of CP-conserving new physics is not significantly large, the constraints 
obtained by us on CP-odd operators are in a way the most conservative estimates of the allowed parameter space. 
Of course, the limits from EDM analyses remain the same in all the cases.

\item

A comparison between the relative strengths of the EDM and LHC constraints, we emphasize, is most 
transparent only when the CP-violating operators alone are considered, since the CP-conserving ones
have no role in EDMs. Their inclusion, albeit via marginalization in the LHC global fits, will serve 
to relax the corresponding limits beyond what we have obtained. But then, it ceases to be a one-to-one
comparison between the two kinds of constraints, something that we have intended to do from the beginning.

\end{itemize}

\begin{table}[t]
 \begin{center}
\begin{tabular}{|c|c|c|c|c|}
  \hline
%    &  & &\\
  {\bf Couplings}  & \multicolumn{2}{|c|}{\bf LHC data}& \multicolumn{2}{|c|}{\bf EDM}\\
  \cline{2-5}
  &{2P case} &{3P case} &{2P case} &{3P case} \\
  \hline
    & & & &\\
  $|C_{WWh}|$               & $0~-~60$   &  $0~-~60$    & $ 0~-~0.17 $ & $0~-~55$  \\
    & & & &\\
  \hline 
   & & & &\\
  $|C_{ZZh}|$               & $25~-~100$ & $25~-~80$      & $0.11~-~0.20$ & $0.15~-~33$  \\
    & & & &\\
    \hline    
   & & & &\\
  $|C_{\gamma\gamma h}|$    & $0~-~0.8$  & $0~-~0.5$       & $0~-~0.16$ & $0.02~-~52$ \\
    & & & &\\
  \hline
   & & & &\\
  $|C_{\gamma zh}|$         & $20~-~25$  & $15~-~25$      & $0.03~-~0.25$ & $0.05~-~110$ \\
    & & & &\\
  \hline 
   & & & &\\ 
  $|C_{WW\gamma}|$          & $0~-~40$   & $15~-~40$       & $0~-~0.15$ &  $0.02~-~47$ \\
    & & & &\\
  \hline 
   \end{tabular}
  \end{center}
  \caption{Limits on CP-odd coupling strengths from LHC data and EDM measurements for $\Lambda$=1 TeV.  
  2P and 3P stand for two parameter nonzero and three parameter nonzero cases respectively.}
  \label{tab:coupling strengths}
  \end{table}

% The observations mentioned above bring to notice the general feature of correlations existing
% among the allowed values of the CP-violating parameters entering the Effective Lagrangian description.
% The allowed values and the sign relationships that these five parameters bear to each other is 
% indicative of the way they contribute to the various $VVH$ and triple gauge boson couplings. 
% The parameters contributing to the CP-violating vertices at tree level, which do not have their
% SM counterparts are severely constrained. This is exemplified by the parameters, 

We have already seen that the Lorentz structure of anomalous CP-odd couplings is unique  and 
these are just some linear combinations of the CP-odd parameters (see Table~\ref{tab:Anomalous_vertices}). 
Since in all the observables these couplings enter directly,
it is instructive to know the kind of values these 
couplings can take as a result of our analysis presented above. 
In the calculation of $S,T$ and $U$ parameters 
all the CP-odd couplings directly enter. In the global analysis only $C_{VVh}$ couplings 
participate, therefore, constraints on $C_{WWV}$ from the LHC data are indirect.  
In EDM calculations only $C_{\gamma\gamma h}, C_{\gamma Z h}$ and $C_{WW\gamma}$ enter directly and therefore 
limits obtained on $C_{WWh}$ and $C_{ZZh}$ are also indirect.
The limits on the strengths of the anomalous couplings are listed in Table \ref{tab:coupling strengths}. 
Since electroweak precision bounds on $\tilde f_i$ are the weakest~\footnote{Although the EWP 
constraints are weaker in general, we find that the limits on $C_{WWV}$ from EWP and LHC data are comparable in 2P case.}, 
in the table we compare bounds on couplings 
due to LHC data and EDMs.  
The comparison is presented for both the two parameter (2P) and three parameter (3P) 
nonzero cases. Since $C_{WWZ}$ is proportional to $C_{WW\gamma}$, the limits on $C_{WW\gamma}$ can be 
easily translated to limits on $C_{WWZ}$. Looking at the LHC limits on the couplings we find that 
$C_{\gamma\gamma h}$ is the most constrained coupling. Also, the LHC limits in 2P and 3P cases are 
comparable. On the other hand the EDM limits on couplings in 2P case is always stronger than in 3P case. 
In 3P case the lower limits result from the three parameter sets $\{\tilde f_B, \tilde f_{BB},\tilde f_{BW}\}$ and 
$\{\tilde f_W, \tilde f_{WW},\tilde f_{BW}\}$. The parameters of these sets are found to be very fine tuned.
It is important to recall that in 3P EDM case, the parameters are varied in the range between -200 to 
200. We find that as we increase this range, the maximum allowed values for couplings also increase. 
We would also like to point out that the correlations among CP-odd couplings are mostly similar to those found 
among CP-odd parameters.
 
The triple gauge boson couplings (TGCs) $WW\gamma$ and $WWZ$ can also be constrained using the collider 
data on gauge boson pair production. Data from Tevatron and LHC are used mainly to constrain the CP-even 
anomalous couplings as the observables used are not sensitive to CP-odd couplings~\cite{Diehl:1998cb,Abazov:2011rk,Aad:2012twa,ATLAS:2012mec,Chatrchyan:2013yaa,Chatrchyan:2013fya}. 
On the other hand, the experimental analyses at LEP which studied the angular distribution of final state particles are sensitive
to CP-odd TGCs. A comparison between the CP-odd sector of TGC 
Lagrangian~\cite{Hagiwara:1986vm,Dawson:2013owa,Gavela:2014vra} and our 
effective Lagrangian (Eq.~\ref{eq:L-bsm}) implies
\begin{eqnarray}
 \frac{C_{WW\gamma}}{\Lambda^2} &=& \frac{s_W}{m_W^2}~ \tilde \kappa_\gamma \\
 \frac{C_{WWZ}}{\Lambda^2} &=& \frac{c_W}{m_W^2}~ \tilde \kappa_Z.
\end{eqnarray}
Using the relation between $C_{WW\gamma}$ and $C_{WWZ}$ we get, 
 $\tilde \kappa_Z = -t_W^2~\tilde \kappa_\gamma$.
At 68\% CL, the combined LEP limits on $\tilde \kappa_Z$ are~\cite{Beringer:1900zz}
\begin{equation}
-0.14 \le \tilde \kappa_Z \le -0.06. 
\end{equation}
These limits when translated on $C_{WWZ}$ and $C_{WW\gamma}$ become, 
\begin{eqnarray}
 -0.19 &\le& \frac{C_{WWZ}}{\Lambda^2}~[{\rm TeV}^{-2}] \le -0.08, \nn \\
 0.15  &\le& \frac{C_{WW\gamma}}{\Lambda^2}~[{\rm TeV}^{-2}] \le 0.36.
\end{eqnarray}
Note that these limits are comparable to the limits obtained from EDMs in 2P case, however, 
information on the sign of the couplings is also available.\\
{ Other than the $VVh$ vertices considered in our analysis, quartic $VVhh$ vertices also arise out of
gauge invariant CP violating operators. One can thus expect some correlated phenomenology from the trilinear
and quartic interactions, since the former arise essentially on replacing one Higgs by its vacuum expectation value
in the quartic terms. In the analysis presented in this work, the production and decay channels considered are
not affected at tree level by such quartic $VVhh$ vertices. Thus in the context of the present analysis, any constraints
on such quartic vertices from observed data are likely to be weaker than those obtained from the $VVh (V = W ,Z,\gamma)$ 
effective interactions, since the quartic couplings would entail Higgs pair production. For example, the $VVhh$ vertex may 
contribute in addition to the $hhh$ vertex toward a di-Higgs final state.
However, one needs to wait for a large volume of data on Higgs pair production to see such 
correlated phenomena. In general, limits stronger than what we have obtained are not expected. 
Also, the contributions from such $VVhh$ CP-odd vertices to EDMs and EWP observables come at higher loop
levels and thus are expected to be substantially weaker than what we have obtained for the trilinear terms.
}

{ It is a well-known fact that the observed baryon asymmetry in our 
universe cannot be explained by just the CP-violating phase of the  Cabibbo-Kobayashi-Maskawa
(CKM) matrix in the SM. The presence of additional sources of CP-violating 
operators arising from the anomalous $VVh$ interactions may in principle 
explain the observed baryon asymmetry of the universe. However, a more
careful scrutiny of this picture reveals that our CP-violating operators are not sufficient
to trigger strongly the first order electroweak phase transition 
required for the baryogenesis. For this, one has to extend the Higgs
sector of the SM by introducing new particles which couple to the
Higgs boson and thus modify the Higgs potential such that it leads
to a strongly first order electroweak phase transition \cite{Morrissey:2012db}.}

\section{Summary and Conclusions}
We have analyzed CP-odd  { $VVh(V = W, Z, \gamma)$ and $WWV(V = Z, \gamma)$}  interactions
in terms of gauge invariant dimension-6 operators, obtained as the artifacts of physics beyond
the standard model. The most complete set, comprising five gauge-Higgs operators, has been
taken into account. We have derived constraints on the coefficients of such operators
using  electroweak precision data, LHC data on Higgs and limits on the electric dipole 
moments of the neutron and the electron. With $\Lambda$ as the scale suppressing the CP-violating 
operators, precision parameters as well as LHC observables receive contributions $\sim 1/\Lambda^4$ from
the CP-odd couplings, while contributions $\sim 1/\Lambda^2$ to EDM observables are expected.
The constraints obtained from the $S, T$ and $U$ parameters are the weakest, while the bounds from EDMs 
are the strongest with two nonvanishing operators. The global analysis of Higgs data from the LHC 
puts stronger constraints on those CP-odd effective couplings which contribute to $h \rightarrow \gamma\gamma$,
as compared to those which do not. We also indicate situations where large values of certain couplings are allowed
by all constraints, when they appear in combination. 
The constraints coming from LEP on CP-odd form factors $C_{WW\gamma}$ and $C_{WWZ}$ are consistent with our 
limits obtained from EDMs in the case when any two out of five parameters are nonzero. 
It may be of interest to find out new physics scenarios where,
by integrating out heavy degrees of freedom, one may arrive at large correlated values of such operators.
% (\textbf{\textit{Their implications for the high-energy run of the LHC are under investigation, and will be reported in a 
% subsequent work~\cite{DGMS:2015xx}}})
{ In a subsequent work, ~\cite{DGMS:2015xx} we hope to discuss some observables that may help one in probing these operators in the 13 
 and 14 TeV { LHC} runs.
}

% \end{itemize}

\section*{Acknowledgements}
We thank Shankha Banerjee, Jyotiranjan Beuria, Nabarun Chakraborty, Arghya Choudhury, Anushree Ghosh and 
Tanumoy Mandal for fruitful discussions. 
The work of S.D., B.M. and A.S. is partially supported by funding available from the Department
of Atomic Energy, Government of India, for the Regional Centre for Accelerator-based Particle
Physics (RECAPP), Harish-Chandra Research Institute. D.K.G. would like to acknowledge the
hospitality of RECAPP at the initial stage of this project. B.M. acknowledges the hospitality 
of Indian Association for the Cultivation of Science,  Kolkata. A.S.
would like to thank CP3-Louvain, Belgium for hospitality while part of the work was carried
out.

% \newpage
% \appendix
\begin{appendices} 
\section{Gauge boson two-point functions in presence of CP-odd couplings} \label{App:AppendixA}

We define the two-point function for electroweak gauge bosons $V_1$ and $V_2$ as,
\begin{eqnarray}
\Pi_{V_1V_2}^{\mu\nu}(p^2) = g^{\mu\nu}\Pi_{V_1V_2}(p^2) + p^\mu p^\nu{\tilde \Pi}_{V_1V_2}(p^2). 
\end{eqnarray}
In electroweak precision observables only $\Pi_{V_1V_2}(p^2)$ contribute. Below we give their
expressions due to CP-odd couplings in terms of one-loop scalar functions $A_0$ and $B_0$. 

 \begin{align}
  \Pi_{\gamma\gamma}(p^2) &=
  \frac{g^2 m_W^2}{18 \Lambda^4} \left(3 p^2 \left(C_{\gamma\gamma h}^2 \left(2 m_h^2-p^2\right) B_0(p^2,0,m_h^2)
  -2 C_{WW\gamma}^2 \left(2 m_W^2+p^2\right) B_0(p^2,m_W^2,m_W^2) \right.\right.\nn \\
   & \left.+~C_{\gamma Z h}^2 A_0(m_Z^2)+4 C_{WW\gamma}^2 A_0(m_W^2)\right)-3 C_{\gamma\gamma h}^2 m_h^4 B_0(p^2,0,m_h^2) 
    -3 C_{\gamma Z h}^2 \left(m_h^4-2 m_h^2 \left(m_Z^2+p^2\right) \right. \nn \\
    &\left. +~\left(m_Z^2-p^2\right)^2\right) B_0(p^2,m_Z^2,m_h^2) + 3 A_0(m_h^2) \left(C_{\gamma\gamma h}^2
   \left(m_h^2+p^2\right)+C_{\gamma Z h}^2 \left(m_h^2-m_Z^2+p^2\right)\right) \nn \\
    & +~3 C_{\gamma Z h}^2 A_0(m_Z^2) \left(m_Z^2-m_h^2\right)+p^2 \left(7 C_{\gamma\gamma h}^2
   \left(p^2-3 m_h^2\right) +~7 C_{\gamma Z h}^2 \left(p^2-3 \left(m_h^2+m_Z^2\right)\right) \right. \nn \\
   & \left.\left. +~2 C_{WW\gamma}^2 \left(12 m_W^2+7 p^2\right)\right)\right),
   \end{align}

   \begin{align}
  \Pi_{\gamma Z}(p^2) &=
\frac{g^2 m_W^2}{18 \Lambda^4} \left(-3 C_{\gamma\gamma h} C_{\gamma Z h} \left(m_h^2-p^2\right)^2 B_0(p^2,0,m_h^2)
   -3 C_{\gamma Z h} C_{ZZh} \left(m_h^4-2 m_h^2 \left(m_Z^2+p^2\right) \right.\right. \nn \\
   & \left. +~\left(m_Z^2-p^2\right)^2\right) B_0(p^2,m_Z^2,m_h^2)-6 C_{WW\gamma} C_{WWZ} p^2 \left(2 m_W^2+p^2\right)
   B_0(p^2,m_W^2,m_W^2) \nn \\
   & +~3 C_{\gamma Z h} A_0(m_h^2) \left(C_{\gamma\gamma h} \left(m_h^2+p^2\right)+C_{ZZh} \left(m_h^2-m_Z^2+p^2\right)\right)+3 C_{\gamma Z h}
   C_{ZZh} A_0(m_Z^2) \nn \\
   & \left(-m_h^2+m_Z^2+p^2\right)+12 C_{WW\gamma} C_{WWZ} p^2 A_0(m_W^2)+p^2 \left(7 p^2 (C_{\gamma Z h} (C_{\gamma\gamma h}+C_{ZZh})\right. \nn \\
   & \left.\left. +~2 C_{WW\gamma} C_{WWZ})-21 C_{\gamma Z h} \left(m_h^2 (C_{\gamma\gamma h}+C_{ZZh})+C_{ZZh} m_Z^2\right)+24 C_{WW\gamma} 
   C_{WWZ} m_W^2\right)\right),
   \end{align}

    \begin{align}
   \Pi_{Z Z}(p^2) &=
 \frac{g^2 m_W^2}{18 \Lambda^4} \left(3 p^2 \left(C_{\gamma Z h}^2 \left(2 m_h^2-p^2\right) B_0(p^2,0,m_h^2)-2 C_{WWZ}^2 \left(2 m_W^2+p^2\right)
    B_0(p^2,m_W^2,m_W^2) \right.\right. \nn \\
    & \left. +~ 4 C_{WWZ}^2 A_0(m_W^2)+C_{ZZh}^2 A_0(m_Z^2)\right)-3 C_{\gamma Z h}^2 m_h^4 B_0(p^2,0,m_h^2)-3 C_{ZZh}^2 \left(m_h^4-2
    m_h^2 \left(m_Z^2+p^2\right) \right. \nn \\
    & \left. +~\left(m_Z^2-p^2\right)^2\right) B_0(p^2,m_Z^2,m_h^2)+3 A_0(m_h^2) \left(C_{\gamma Z h}^2
    \left(m_h^2+p^2\right)+C_{ZZh}^2 \left(m_h^2-m_Z^2+p^2\right)\right) \nn \\
    & +~ 3 C_{ZZh}^2 A_0(m_Z^2) \left(m_Z^2-m_h^2\right)+p^2 \left(7 C_{\gamma Z h}^2
    \left(p^2-3 m_h^2\right)+2 C_{WWZ}^2 \left(12 m_W^2+7 p^2\right) \right. \nn \\
    & \left.\left. +~7 C_{ZZh}^2 \left(p^2-3 \left(m_h^2+m_Z^2\right)\right)\right)\right),
    \end{align}

       \begin{align}
   \Pi_{WW}(p^2) &=
\frac{g^2 m_W^2}{18 \Lambda^4 p^2} \left(3 \left(-C_{WW\gamma}^2 \left(m_W^2-p^2\right)^2 \left(m_W^2+p^2\right) B_0(p^2,m_W^2,0)
     +C_{WWh}^2 (-p^2) \left(m_h^4 \right.\right.\right. \nn \\
     & \left. -~2 m_h^2 \left(m_W^2+p^2\right)+\left(m_W^2-p^2\right)^2\right) B_0(p^2,m_W^2,m_h^2)-C_{WWZ}^2 \left(m_W^6-m_W^4 \left(2 m_Z^2+p^2\right)\right. \nn \\
   & \left. +~m_W^2 \left(m_Z^4+8 m_Z^2 p^2-(p^2)^2\right)+p^2 \left(m_Z^2-p^2\right)^2\right) B_0(p^2,m_W^2,m_Z^2)-C_{WWZ}^2 A_0(m_Z^2)
   \left(m_W^2+p^2\right) \nn \\
   & \left. \left(m_W^2-m_Z^2-p^2\right)\right)+3 A_0(m_W^2) \left(C_{WW\gamma}^2 \left(m_W^4-10 m_W^2 p^2+(p^2)^2\right)
    +~ C_{WWh}^2 p^2 \left(-m_h^2+m_W^2+p^2\right) \right.  \nn \\
    & \left. +~C_{WWZ}^2 \left(m_W^4-m_W^2 \left(m_Z^2+10 p^2\right)+p^2 \left(p^2-m_Z^2\right)\right)\right)+3 C_{WWh}^2
   p^2 A_0(m_h^2) \left(m_h^2-m_W^2+p^2\right) \nn \\
   & +p^2 \left(C_{WW\gamma}^2 \left(87 m_W^4-14 m_W^2 p^2 +~ 7 (p^2)^2\right)+7 C_{WWh}^2 p^2
   \left(p^2-3 \left(m_h^2+m_W^2\right)\right)+C_{WWZ}^2 \right. \nn \\
   &\left.\left. \left(-7 p^2 \left(2 m_W^2+3 m_Z^2\right)+87 m_W^2 \left(m_W^2+m_Z^2\right)+7
   (p^2)^2\right)\right)\right). \\ \nn
    \end{align}
   
   Out of these, $\Pi_{\gamma\gamma}, \Pi_{\gamma Z}$ and $\Pi_{ZZ}$ vanish at $p^2=0$. Note that in $\Pi_{WW}$ 
   there is an overall $1/p^2$ dependence. We would like to mention that both $\Pi_{WW}$ and its derivative 
   converge smoothly in $p^2\to 0$ limit.
   The one-loop 
   scalar functions in $n=4-2\epsilon$ dimensions are given by, 
   \begin{eqnarray}
    A_0(m_0^2) &=& \int \frac{d^nl}{(2\pi)^n}~~ \frac{1}{l^2-m_0^2} \nn \\
               &\equiv& \frac{1}{16 \pi^2} m_0^2 \left[ \frac{1}{\epsilon} +1 -{\rm ln}(m_0^2) \right], \\
     B_0(p^2, m_0^2,m_1^2) &=& \int \frac{d^nl}{(2\pi)^n}~~ \frac{1}{(l^2-m_0^2)~ ((l+p)^2-m_1^2)} \nn \\
     &\equiv& \frac{1}{16 \pi^2} \left[ \frac{1}{\epsilon} - \Delta(p^2,m_0^2,m_1^2) \right],
   \end{eqnarray}

   where, 
   \begin{eqnarray}
    \Delta(p^2,m_0^2,m_1^2) = \int_0^1 dx~~ {\rm ln}\left[-x (1 - x) p^2 + x (m_1^2 - m_0^2) + m_0^2 \right].
   \end{eqnarray}
This form is suitable for computing $B_0$ and its derivative with respect to $p^2$ at $p^2=0$, which 
we require to calculate $S, T$ and $U$ parameters discussed in section 3.
   
   \end{appendices}
   
% % % % % % % % % % %    

\end{document}